\documentclass{article}

\usepackage[utf8]{inputenc} % allow utf-8 input
\usepackage[T1]{fontenc}    % use 8-bit T1 fonts
\usepackage{hyperref}       % hyperlinks
\usepackage{url}            % simple URL typesetting
\usepackage{booktabs}       % professional-quality tables
\usepackage{amsfonts}       % blackboard math symbols
\usepackage{nicefrac}       % compact symbols for 1/2, etc.
\usepackage{microtype}      % microtypography
\usepackage{lipsum}
\usepackage{fancyhdr}       % header
\usepackage{graphicx}       % graphics
\graphicspath{{media/}}     % organize your images and other figures under media/ folder

\usepackage{amsmath}
\usepackage{amssymb}

\begin{document}

\title{Space-resolved stress correlations and viscoelastic moduli for 
polydisperse systems:  the faces of the stress noise} 

\author{J\"org Baschnagel, Alexander N. Semenov$^{*}$ \\
Institut Charles Sadron, CNRS - UPR 22, Universit\'e de Strasbourg,\\
23 rue du Loess, BP 84047, 67034 Strasbourg Cedex 2, France}

\date{\today} 
\maketitle

\begin{abstract}

Several advances in the theory of space-resolved viscoelasticity of 
liquids and other amorphous systems are discussed in the present 
paper.  In particular, considering long-time regimes of stress 
relaxation in liquids we obtain the generalized compressibility 
equation valid for systems with mass polydispersity, and derive a 
new relation allowing to calculate the wavevector-dependent 
equilibrium transverse modulus in terms of the generalized structure 
factors.  Turning to the basic relations between the 
spatially-resolved relaxation moduli and the spatio-temporal 
correlation functions of the stress tensor, we provide their new 
derivation  based on a conceptually simple argument that does not 
involve consideration of non-stationary processes.  We also elucidate 
the relationship between the stress noise associated with the 
classical Newtonian dynamics and the reduced deviatoric stress 
coming from the Zwanzig-Mori projection operator formalism.  The 
general relations between the stress noise and the tensor of 
relaxation moduli are discussed as well.

\end{abstract}

\newpage

\section{Introduction}

Mechanical stress and {the related\/} viscoelastic moduli are among 
the main rheological characteristics defining the flow properties of 
complex fluids including polymer melts and solutions~\cite{RubColby}, 
molten metallic alloys, soft-matter systems and glass-forming 
(supercooled) liquids.~\cite{Bird,Heuer,Ferry,Nicolas,Fuchs,Berthier} 
Flow behavior of such systems combine the properties of liquids and 
elastic solids.  These complex systems are typically heterogeneous 
(being amorphous in nature) and {are\/} characterized in the general 
case by the stress tensor fields, $\sigma_{\alpha\beta}(\underline r,t)$, and the associated 
relaxation functions resolved not only in time $t$, but also in space 
({here} $\alpha ,\beta$ {are } the Cartesian components).  {The relevant 
examples} include the generalized (wavevector, $\underline q$, dependent) material 
functions like the shear, $G(q,t)$, and longitudinal, $L(q,t)$, relaxation 
moduli~\cite{JCP164505} which show a spectrum of relaxation times 
and correlation lengths characterizing amorphous systems and defining 
their micro- and nano-rheological properties.~\cite{microR} 

{The whole} set of such {$\underline q$-dependent relaxation} functions
defines the tensor $E_{\alpha\beta\gamma\delta}(\underline q,t)$ of {generalized} relaxation
moduli~\cite{Polymers2024,SM2025}.  The long-time limit of the 
elasticity tensor, $E_{\alpha\beta\gamma\delta}^e(\underline q)\equiv E_{\alpha\beta
\gamma\delta}(\underline q,t\to\infty )$, characterizes the purely 
static (equilibrium) response of the system to a deformation.  As 
demonstrated in ref.~\cite{Polymers2024}, the equilibrium elasticity 
tensor in fluid regime can be characterized by just two basic moduli:  
$L_e(q)$ defining the longitudinal stress response to a weak longitudinal 
deformation wave, and $M_e(q)$ defining the transverse stress response 
to the same deformation.  

An efficient way to obtain the relaxation moduli by simulations in 
the bulk limit (ie at $q=0$) is to use the stress-fluctuation 
relations~\cite{SoftM7867} expressing them in terms of 
time-dependent correlation functions.  However, this approach is hard 
to implement at $q>0$~\cite{JCP164505} and, moreover, it seems to 
require very long simulation runs in the case $t\to\infty$.  To overcome 
these difficulties in obtaining $L_e(q)$ and $M_e(q)$ we derive in 
section~\ref{sII} a number of relations providing these moduli in 
terms of correlation functions of structural variables taken at the 
same time thus {avoiding the need of long runs}.

The generalized viscoelastic moduli can be obtained using the 
stress-fluctuation relations generalized to the case of a finite 
wavevector $\underline q$.~\cite{SM2018,Polymers2024} Some of these relations 
(involving $G(q,t)$ and $L(q,t)$) are known for a long 
time~\cite{EvansMor}, while the equation for the transverse modulus 
$M(q,t)$ was first obtained in ref.~\cite{SM2018}.  

{A general tensorial relation between} the tensor $C_{\alpha\beta\gamma\delta}(\underline 
q,t)$ of 
spatio-temporal correlation functions of the stress field and the 
elasticity tensor $E_{\alpha\beta\gamma\delta}(\underline q,t)$ was obtained in ref.~\cite{Polymers2024} 
(cf eqn (115) there) using the fluctuation-dissipation theorem (FDT) 
and the concept of stress noise.~\cite{SM2018,JCP164505,Polymers2024} 
{However, this general relation was derived based on} consideration 
of the dynamics implying that the flow (at the wavevector $\underline q$) is 
arrested at $t<0$, but is released later.  To validate this important 
relation between $C$ and $E$ tensors (quoted in eqn~\ref{e2}) we present 
its novel derivation (see section~\ref{sIII}) which {avoids imposing} 
the no-flow constraint.  

It was recently established~\cite{SM2025} that the memory kernel 
$M_{\alpha\beta\gamma\delta}(\underline q,t)$ of the Zwanzig-Mori (ZM) projection operator 
formalism~\cite{FuchsPRL2017,FuchsJCP2018,SM2025} can be identified 
with the tensorial correlation function $C^n_{\alpha\beta\gamma\delta}(\underline 
q,t)$ of the stress noise, 
$\sigma_{\alpha\beta}^n(\underline q,t)$, {which is present even if a macroscopic flow in the system 
is suppressed} (cf section~\ref{ECn}):  
\begin{equation}M_{\alpha\beta\gamma\delta}(\underline q,t)=C^n_{\alpha\beta\gamma
\delta}(\underline q,t)\label{MCn}\end{equation}
{At the ensemble-averaged level, eqn~\ref{MCn} reveals a close 
relationship between the projected dynamics based on the ZM 
approach~\cite{SM2025} and the classical dynamics of stress noise 
considered in ref.~\cite{Polymers2024}.  This result begs a natural 
question on how the underlying processes---the projected dynamics of 
the reduced deviatoric stress~\cite{SM2025} and the classical dynamics 
of the stress noise---are related to each other.  This question is 
addressed in section~\ref{sI} leading to important conclusions 
regarding both dynamics (see section~\ref{sI3}).} 

The scope of the paper is:  {In the next section we
introduce the basic dynamical variables and the classical general 
relationships between them.  The $\underline q$-dependent elastic moduli in the 
static regime, $t\to\infty$, are considered in section~\ref{sII}; the general 
relations between these moduli and the structural correlation 
functions are established there.  In section~\ref{sIII} we present a 
new derivation of the general relation (cf eqn~\ref{e2}) between the 
stress-correlation tensor $C_{\alpha\beta\gamma\delta}(\underline q,t)$ and the elasticity tensor 
$E_{\alpha\beta\gamma\delta}(\underline q,t)$.  The relationship between the reduced deviatoric stress, 
$\sigma^{rd}(\underline q,t)$, of the projected dynamics and the classical dynamics of 
stress noise, $\sigma_{\alpha\beta}^n(\underline q,t)$, is elucidated in section~\ref{sI}.  A discussion 
and summary of the main results are provided in section~\ref{DS}.} 

\section{The basic equations and definitions of the fields encoding 
liquid dynamics}

\subsection{Density, velocity and stress fields, and the generalized 
relaxation moduli} 

Let us consider a liquid of $N\gg 1$ particles in volume $V$.  A 
particle `$i$' has mass $m_i$, position $\underline r_i$ and velocity $\underline 
v_i$.  The mean 
particle mass, concentration and mass 
density, respectively, are:
\begin{equation}\bar {m}=(1/N)\sum_im_i,~~c_0=N/V,~\rho_0=\bar {m}c_0\label{mcrho}\end{equation}
The mass density field is defined as
\begin{equation}\rho (\underline r,t)=\sum_im_i\delta (\underline r-\underline r_
i(t))\label{rhort}\end{equation}
while the momentum density is
\begin{equation}\underline J(\underline r,t)=\sum_im_i\underline v_i(t)\delta (\underline 
r-\underline r_i(t))\label{Jrt}\end{equation}
The two fields are related by the mass balance 
equation~\cite{HansDon,BalZop}:
\begin{equation}\frac {\partial\rho (\underline r,t)}{\partial t}=-\frac {\partial}{
\partial r_{\alpha}}J_{\alpha}(\underline r,t)\label{mbal}\end{equation}
where summation over repeated indices is assumed.

Next, we define the concentration and velocity fields following 
ref.~\cite{Polymers2024} where the contribution of each particle is 
weighted with the factor $m_i/\bar {m}$ proportional to the particle mass:  
\begin{equation}c(\underline r,t)=\sum_i\left(m_i/\bar m\right)\delta (\underline 
r-\underline r_i(t))=\rho (\underline r,t)/\bar {m}\label{crt}\end{equation}
\begin{equation}\underline v(\underline r,t)=\frac 1{c_0}\sum_i\left(m_i/\bar m\right
)\underline v_i(t)\delta (\underline r-\underline r_i(t))=\underline J(\underline 
r,t)/\rho_0\label{vrt}\end{equation}
Note that thus defined $c(\underline r,t)$ and $\underline v(\underline r,t)$ are proportional to the mass 
density and the momentum density, respectively.   The main reason 
why the $c$ and $\underline v$ fields are defined with mass factors, $m_i/\bar {
m}$, is that 
with such definitions they are connected to each other and to the 
stress field (cf eqn~\ref{sigr}) by the fundamental conservation laws 
(cf eqn~\ref{mbal} above and eqn~\ref{momeq} below).  Other reasons 
supporting the adopted definitions, eqn~\ref{crt},~\ref{vrt}, are 
considered in detail in section~\ref{DS}.  Obviously, in monodisperse 
systems (with particles of equal mass) eqn~\ref{crt},~\ref{vrt} coincide with 
`conventional' definitions of concentration and collective velocity 
fields~\cite{HansDon,BalZop}.  
 
Their Fourier transforms (indicated solely by the argument $\underline q$) are (cf 
eqn (30) of ref.~\cite{Polymers2024}) 
\begin{equation}c(\underline q,t)\equiv\frac 1V\int c(\underline r,t)e^{-{\rm i}\underline 
q\cdot\underline r}{\rm d}^dr=\frac 1V\sum_i\left(m_i/\bar m\right)\exp(-{\rm i}\underline 
q\cdot\underline r_i)\label{cqt}\end{equation}
\begin{equation}\underline v(\underline q,t)=\frac 1N\sum_i\left(m_i/\bar m\right
)\underline v_i(t)\exp(-{\rm i}\underline q\cdot\underline r_i)\label{vqt}\end{equation}
where $\underline q$ is the wavevector and $d$ is the space dimension.  
The mass current density in Fourier space is
\begin{equation}\underline J(\underline q,t)=\frac 1V\sum_im_i\underline v_i(t)\exp
(-{\rm i}\underline q\cdot\underline r_i)\label{Jqt}\end{equation}
Obviously $\underline J(\underline q,t)=\rho_0\underline v(\underline q,t)$.  Note that the first equation in 
eq.~\ref{cqt} is the definition of the Fourier transform adopted in the 
present paper.  It ensures that physical dimensions of transformed 
and original functions are the same.  

The mass conservation law (cf eq.~\ref{mbal}) in terms of the 
concentration (defined in eqn~\ref{cqt}) and velocity 
(eqn~\ref{vqt}) fields becomes 
\begin{equation}\frac 1{c_0}\frac {\partial c(\underline q,t)}{\partial t}=-{\rm i}\underline 
q\cdot\underline v(\underline q,t)\label{mqbal}\end{equation}
Furthermore, the fundamental momentum balance equation defines the 
rate of change of momentum density $\underline J(\underline r,t)$:  
\begin{equation}\frac {\partial J_{\alpha}(\underline r,t)}{\partial t}=\frac {\partial}{
\partial r_{\beta}}\sigma_{\alpha\beta}(\underline r,t)\label{momeq}\end{equation}
Here $\sigma_{\alpha\beta}(\underline r,t)$ is the stress tensor field, whose microscopic definition  
is~\cite{Gold2002,Shi2023}
\begin{equation}\sigma_{\alpha\beta}(\underline r,t)=\sum_{i>j}^NF_{ij\alpha}r_{
ij\beta}\int_0^1\delta\left(\underline r-\underline r_i+s\underline r_{ij}\right
){\rm d}s-\sum_im_iv_{i\alpha}v_{i\beta}\delta\left(\underline r-\underline r_i\right
)\label{sigr}\end{equation}
where $\delta$ is the $d$-dimensional delta-function, $F_{ij\alpha}$ is the $\alpha$-component
of the force acting on particle $i$ from particle $j$, $r_{ij\beta}$ is $\beta$-component 
of vector $\underline r_{ij}=\underline r_j-\underline r_i$, and $v_{i\alpha}$ is component $
\alpha$ of the velocity of 
particle $i$ (all vectors here are taken at time $t$).  

Doing Fourier transform of eq.~\ref{sigr} we get the wavevector 
dependent stress (cf refs~\cite{Polymers2024,SM2025,BalZop}):  
\[\sigma_{\alpha\beta}(\underline q,t)\equiv\frac 1V\int\sigma_{\alpha\beta}(\underline 
r,t)e^{-{\rm i}\underline q\cdot\underline r}\,{\rm d}^dr=\]
\begin{equation}=\frac 1V\sum_{i>j}F_{ij\alpha}r_{ij\beta}\left(e^{-{\rm i}\underline 
q\cdot\underline r_i}-e^{-{\rm i}\underline q\cdot\underline r_j}\right)/\left({\rm i}\underline 
q\cdot\underline r_{ij}\right)-\frac 1V\sum_im_iv_{i\alpha}v_{i\beta}e^{-{\rm i}\underline 
q\cdot\underline r_i}\label{G2}\end{equation}
where the first sum includes all disordered pairs $i,j$ of interacting 
particles, and the force $F_{ij\alpha}\equiv\left(\frac {r_{\alpha}}ru'_{ij}(r)\right
)_{\underline r=\underline r_j-\underline r_i}$with $u_{ij}(r)$ being 
the interaction energy of particles $i$ and $j$, and $u'_{ij}(r)\equiv\frac {{\rm d}
u_{ij}(r)}{{\rm d}r}$.  

The momentum equation in the wavevector representation comes 
from eq.~\ref{momeq}:  
\begin{equation}\rho_0\frac {\partial v_{\alpha}(\underline q,t)}{\partial t}={\rm i}
q_{\beta}\sigma_{\alpha\beta}(\underline q,t)\label{e11}\end{equation}
where $v_{\alpha}(\underline q,t)$ is the $\alpha$-component of the $\underline 
q$-dependent velocity field 
(cf eqn~\ref{vqt}).  Note that here and in what follows we always 
assume $\underline q\neq 0$.  

We are now in a position to define the $\underline q$-dependent relaxation moduli 
$E_{\alpha\beta\gamma\delta}(\underline q,t)$.  To this end let us consider an equilibrated amorphous 
(liquid or solid) system which is slightly perturbed by application of 
a weak external force field~\footnote{See eq.~\ref{eqF} in the 
Discussion section.} generating the time-dependent 
strain rate $\dot{\gamma}_{\alpha\beta}(\underline r,t)=\frac {\partial v_{\alpha}
(\underline r,t)}{\partial r_{\beta}}$ whose Fourier transform is 
\begin{equation}\dot{\gamma}_{\alpha\beta}(\underline q,t)={\rm i}q_{\beta}v_{\alpha}
(\underline q,t)\label{gammadot}\end{equation}
The linear response of $\sigma_{\alpha\beta}(\underline q,t)$ (ie the ensemble-averaged change of 
the stress tensor, $\left<\delta\sigma_{\alpha\beta}(\underline q,t)\right>$, in the linear order due to 
the perturbation) is related to the ensemble-averaged strain, 
$\left<\dot\gamma_{\alpha\beta}(\underline q,t)\right>$~\cite{HansDon,Polymers2024}:  
\begin{equation}\left<\delta\sigma_{\alpha\beta}(\underline q,t)\right>=\int_{-\infty}^
tE_{\alpha\beta\gamma\delta}(\underline q,t-t')\left<\dot\gamma_{\gamma\delta}(\underline 
q,t')\right>{\rm d}t'\label{e0}\end{equation}
where the Boltzmann superposition principle is taken into 
account.~\footnote{Note that ensemble-averaging is not needed for 
ergodic systems in the macroscopic limit $V\to\infty$.} The lower limit of 
integration $(-\infty$) implies that the flow may be present in the past, 
while the upper limit $(t$) comes from the casuality principle dictating 
that 
\begin{equation}E_{\alpha\beta\gamma\delta}\left(\underline q,t<0\right)\equiv 0\label{e8}\end{equation}
The kernel $E_{\alpha\beta\gamma\delta}(\underline q,t)$ in eq.~\ref{e0} is the tensor of the generalized 
$(\underline q,t)$-dependent elastic moduli.  Exactly the same elasticity tensor can 
be defined via the (ensemble-averaged) stress response to an instant 
infinitesimal harmonic-canonical strain~(for more details see section 
5.1 in ref.~\cite{Polymers2024} or section 2 of ref.~\cite{SM2018}).  

In the genuinely static regime the elasticity tensor tends to the 
so-called equilibrium elasticity tensor,~\cite{Polymers2024,SM2025} 
\begin{equation}E_{\alpha\beta\alpha'\beta'}^e(\underline q)\equiv E_{\alpha\beta
\alpha'\beta'}(\underline q,t\to\infty )\label{fnew}\end{equation}
Note that $t\to\infty$ in eqn~\ref{fnew} is considered as a very long 
(perhaps inaccessible in practice) time sufficient for a complete 
equilibration of the system.

\subsection{\label{ECn}Stress correlation functions, deterministic 
stress and stress noise} 

The stress correlation function (whose relation to the elasticity 
tensor $E_{\alpha\beta\gamma\delta}(\underline q,t)$ is considered in section~\ref{sIII}) is defined 
as~\cite{SM2018,Polymers2024} 
\begin{equation}C_{\alpha\beta\gamma\delta}(\underline q,t)\equiv\frac VT\left<\sigma_{
\alpha\beta}(\underline q,t+t')\sigma_{\gamma\delta}(-\underline q,t')\right>,~~
q\neq 0\label{e4}\end{equation}
(cf eqn (72) in ref.~\cite{Polymers2024} and eqn (6) in 
ref.~\cite{SM2018}).  Here $T$ is temperature in energy units and 
angular brackets, $\left<...\right>$, mean both ensemble and gliding averaging 
(with respect to $t'$).  {Note that the above equation applies to
equilibrated systems, hence $\left<\sigma_{\alpha\beta}(\underline q,t)\right>=0$ for $
q\neq 0$.  Therefore, the 
instantaneous stress tensor, $\sigma_{\alpha\beta}(\underline q,t)$, itself represents the stress 
fluctuation if $q\neq 0$.} The correlation function defined in eq.~\ref{e4} 
is real and shows both minor and major 
symmetries~\cite{SM2018,SM2025} 
\begin{equation}C_{\alpha\beta\gamma\delta}(\underline q,t)=C_{\beta\alpha\gamma
\delta}(\underline q,t)=C_{\gamma\delta\alpha\beta}(\underline q,t)\label{ne5}\end{equation}

The microscopic stress fields can be generally split in two 
contributions:  the deterministic stress $\sigma_{\alpha\beta}^d(\underline q,t)$ and the stress noise 
$\sigma_{\alpha\beta}^n(\underline q,t)$~\cite{SM2018,Polymers2024}:  
\begin{equation}\sigma_{\alpha\beta}(\underline q,t)=\sigma_{\alpha\beta}^d(\underline 
q,t)+\sigma_{\alpha\beta}^n(\underline q,t)\label{e6}\end{equation}
The deterministic stress is defined by the deformation (flow) history 
(cf eqn~\ref{e0}):  
\begin{equation}\sigma^d_{\alpha\beta}(\underline q,t)=\int_{-\infty}^tE_{\alpha
\beta\gamma\delta}\left(\underline q,t-t'\right){\rm i}v_{\gamma}\left(\underline 
q,t'\right)q_{\delta}{\rm d}t'\label{e7}\end{equation}
where $v_{\gamma}\left(\underline q,t'\right)$ is the $\gamma$-component of Fourier transform of the 
velocity field (cf eqn~\ref{vqt}).  Note also that the fields (like 
$\sigma_{\alpha\beta}(\underline q,t)$, $v_{\gamma}\left(\underline q,t'\right)$, etc) involved in the above equations are 
microscopic, and we never consider coarse-graining either in space or 
in time in the present paper.  Note also that eqs.~\ref{e7} 
and~\ref{e0} take into account that the system is macroscopically 
uniform.

The deterministic stress, $\sigma^d_{\alpha\beta}(\underline q,t)$, is the stress part directly related 
to the prior microscopic velocity field, $v_{\gamma}\left(\underline q,t'\right)$, in the system.  We 
emphasize that no ensemble-averaging is applied to either $v_{\gamma}\left(\underline 
q,t'\right)$ in 
eq.~\ref{e7} or the total stress, $\sigma_{\alpha\beta}(\underline q,t)$, in eq.~\ref{e6}.  Thus, 
$\sigma^d_{\alpha\beta}(\underline q,t)$ can show short-scale fluctuations.  

The stress noise, $\sigma_{\alpha\beta}^n(\underline q,t)$ in eq.~\ref{e6}, is the complementary 
stress part which is due to structural and other thermal fluctuations 
unrelated to the flow or strain at the given wavevector 
$\underline q$.~\cite{SM2018,Polymers2024} Formally $\sigma_{\alpha\beta}^n(\underline 
q,t)$ is defined as the 
total stress devoid the deterministic part.  

While the rhs of eq.~\ref{e7} and eq.~\ref{e0} seem to be equivalent 
in view of eq.~\ref{gammadot}, there is a conceptual difference 
between $\sigma^d_{\alpha\beta}(\underline q,t)$ and $\delta\sigma_{\alpha\beta}
(\underline q,t)$.  It stems from their physical 
interpretation:  $\left<\delta\sigma_{\alpha\beta}(\underline q,t)\right>$ and $\left
<\dot\gamma_{\gamma\delta}(\underline q,t')\right>$ in eq.~\ref{e0} are the 
ensemble-averaged variables characterizing the perturbed ensemble 
(with an `imposed' flow field).  By contrast, both $\sigma^d_{\alpha\beta}(\underline 
q,t)$ and 
$v_{\gamma}\left(\underline q,t'\right)$ in eq.~\ref{e7} characterize an individual trajectory in the 
phase space (of a member of an equilibrium canonical or 
microcanonical ensemble).~\footnote{Note that in order to avoid a 
fictitious divergence of the integral in eq.~\ref{e7}, a factor $e^{\epsilon t}$ 
should be introduced in the integrand and the limit $\epsilon\to 0$ should be 
taken as the last step.} 

The correlation function of stress noise can be defined in analogy 
with eq.~\ref{e4}:  
\begin{equation}C^n_{\alpha\beta\gamma\delta}(\underline q,t)\equiv\frac VT\left
<\sigma^n_{\alpha\beta}(\underline q,t+t')\sigma^n_{\gamma\delta}(-\underline q,
t')\right>\label{e3}\end{equation}
Note that $C^n_{\alpha\beta\gamma\delta}(\underline q,t)$ must vanish at long $t$ since $\left
<\sigma_{\alpha\beta}^n(\underline q,t')\right>=0$ and 
the terminal relaxation time is finite:  $C^n(\underline q,t)\to 0$ at $t\to\infty$.  

Using the fluctuation-dissipation theorem 
(FDT)~\cite{LL5,BalZop,HansDon} it was recently 
shown~\cite{Polymers2024} that the correlation function of the stress 
noise $(C^n$) is closely related to the elasticity kernel $(E$):  
\begin{equation}C_{\alpha\beta\gamma\delta}^n(\underline q,t)=E_{\alpha\beta\gamma
\delta}(\underline q,\left|t\right|)-E_{\alpha\beta\gamma\delta}^e(\underline q)\label{f5}\end{equation}
where $\left|t\right|$ in the rhs is required since $C^n$ is even in time, while the 
convention $E=0$ at $t<0$ is adopted in the present paper.  This 
equation is reminiscent of the classical Einstein's relation connecting 
the mobility of a (tracer) particle in a liquid (defining the particle 
drift velocity in response to a driving force) with its self-diffusion 
constant $D$ reflecting the Brownian motion in the system~\cite{LL6}.  
More precisely, eq.~\ref{f5} is also akin to the relation between the 
particle friction coefficient (inverse mobility) and the correlation 
function of the random force (noise term) in the Langevin 
equation.~\cite{Leq,LL5} Noteworthily, the Einstein's relation was 
generalized in order to predict the mean-square displacement of the 
centre-of-mass of a polymer chain in a melt showing a non-trivial 
time-dependence~\cite{ad1,FaragoPRE051807,SemJCP244905,ad7,ad8}.  

In section~\ref{sIII} the basic equation~\ref{f5} is used as a starting 
point to derive the general equation connecting $C_{\alpha\beta\gamma\delta}(\underline 
q,t)$ and 
$E_{\alpha\beta\gamma\delta}(\underline q,t)$ which, as mentioned in the Introduction, was obtained in a 
less straightforward way in ref.~\cite{Polymers2024}.  

In the next section we present a new approach allowing to obtain the 
static $q$-dependent elastic moduli of a liquid (or a well-equilibrated 
amorphous solid) by simulations.  

\section{\label{sII}Equilibrium elastic moduli of a liquid or amorphous
solid} 

\subsection{The generalized compressibility equation (GCE)}

A deformation (flow) of an amorphous system generates a stress 
response defined by three main $\underline q$-dependent relaxation 
moduli~\cite{SM2018,Polymers2024}:  the shear modulus, 
$G(q,t)=E_{2121}(\underline q,t)$, the longitudinal modulus, $L(q,t)=E_{1111}(\underline 
q,t)$, and the 
mixed (transverse) modulus, $M(q,t)=E_{2211}(\underline q,t)$, {where the
Cartesian components correspond to the} naturally rotated coordinates 
(NRC) with axis 1 parallel to the wavevector $\underline q$~\cite{PRE2023}.  
The {\em static\/} response in liquids is given by the 
long-time limit ($t\to\infty$) of these relaxation 
moduli:~\cite{SM2018,Polymers2024,SM2025} 
\begin{equation}G_e(q)\equiv G(q,t\to\infty )=0,~L_e(q)\equiv L(q,t\to\infty ),~
M_e(q)\equiv M(q,t\to\infty )\label{GLM}\end{equation}
In amorphous glassy systems a complete relaxation is impossible as 
the terminal relaxation time in such systems is not accessible 
experimentally, so one has to distinguish between the long-time 
elastic moduli, $G_{pl}(q)>0$, $L_{pl}(q)$, and $M_{pl}(q)$, corresponding to the 
glassy plateau, and the genuine static moduli, $G_e(q)=0$, $L_e(q)>0$ and 
$M_e(q)>0$, characterizing the stress response upon complete relaxation 
of an amorphous system (which can be achieved in model systems 
using computer simulations, for example, by applying particle swaps 
in polydisperse systems of Lennard-Jones particles~\cite{swaps}).  
That is why we prefer to call $L_e$ and $M_e$ the equilibrium elastic 
moduli in order to distinguish them from the glassy plateaux occurring 
in supercooled liquids at intermediate time-scales which strongly 
increase on cooling towards the glass transition temperature $T_g$ (note 
that the plateau values are often referred to as `non-ergodicity
parameters'~\cite{Goetze,P4,P5,Jan8,Jan10}).  Obviously  the static and
equilibrium moduli are the same in the high-temperature liquid 
regime.  

It is widely accepted that the equilibrium longitudinal modulus $L_e(q)$ 
is related to the common static structure factor $S(q)$, 
\begin{equation}S(q)=\frac 1N\sum_{ij}\left<\exp\left({\rm i}\underline q\cdot\left
(\underline r_i-\underline r_j\right)\right)\right>,\label{eSq}\end{equation}
via the generalized compressibility equation
(GCE)~\cite{HansDon,BalZop,Voronoi,Polymers2024,SM2025}
\begin{equation}L_e(q)=c_0T/S(q),~~q\neq 0\label{IIe4}\end{equation}
For $q\to 0$ 
the above equation reduces to the classical compressibility 
equation~valid for monodisperse systems:~\cite{OZ,HansDon,BalZop} 
\begin{equation}\kappa_T=S(q\to 0)/\left(c_0T\right)\label{kTmono}\end{equation}
where
\begin{equation}\kappa_T=-\left(\frac {\partial P}{\partial\ln V}\right)_T^{-1}=\frac 
1{L_e^{\mbox{\rm bulk}}}\label{IIe5}\end{equation}
is the static isothermal compressibility of the system (here $P$ is the 
mean pressure).  Thus, the longitudinal elastic modulus $L_e(q\to 0)$ 
coincides with the bulk compression modulus $L_e^{\mbox{\rm bulk}}$ of a monodisperse 
liquid.~\cite{PRE042611} However, the latter statement and 
eqn~\ref{kTmono} are not valid for polydisperse systems (as 
discussed in refs.~\cite{HansDon,JCP164505,PRE042611}).  

Moreover, while the GCE, eqn~\ref{IIe4}, is valid for systems with 
particles of equal mass, it is not applicable as such to systems with 
particle mass polydispersity.  A derivation of a more general 
equation, valid also for systems with mass polydispersity, is 
delegated in Appendix B.  This general result can be also 
understood in simpler terms as outlined in the next section.  

\subsection{GCE for polydisperse systems}

Let us consider a liquid at or near the equilibrium.  The free energy 
${\cal F}$ associated with a collective fluctuation of a given amplitude $a$ (like 
a concentration wave at wavevector $\underline q$:~\footnote{Note that by 
concentration in the present paper we always mean the reduced 
density as defined in eqn~\ref{crt},~\ref{cqt}.}  in this case $a=c(\underline q
)$, 
cf eqn~\ref{cqt}) is proportional to the total volume $V$ of the system.  
The typical amplitude $a$ must be therefore small for large systems 
(given that the typical free energy of a thermal fluctuation is 
comparable with the thermal energy $T$), so that ${\cal F}$ can be well 
approximated by a quadratic form, ${\cal F}\simeq\frac {\kappa V}2\left|a\right|^
2$ (where $\kappa$ depends on $q$, 
$\kappa =\kappa (q)$), which also means that the fluctuation statistics are 
Gaussian leading to $\left<\left|a\right|^2\right>=T/(\kappa V)$.~\footnote{Here we take into 
account that $a$ is a complex number with two components (its Re and 
Im parts) involved in two fluctuation waves (with wavevectors $\underline q$ and 
$-\underline q$) whose total free energy is $2{\cal F}$.} Thus, for $a=c(\underline 
q)$ we get 
\[\left<\left|c(\underline q)\right|^2\right>=T/(\kappa V)\]
Defining the generalized structure factor $S_2(q)$ as 
\begin{equation}S_2(q)=\frac 1N\sum_{i,j}\left(m_im_j/\bar m^2\right)\left<\exp\left
({\rm i}\underline q\cdot\left(\underline r_i-\underline r_j\right)\right)\right
>\label{Sqpoly}\end{equation}
(cf eq.~\ref{mcrho}) we find from eqn~\ref{cqcq} 
\[\kappa (q)=\frac T{c_0S_2(q)}\]
Let us now take into account that an increment of the concentration 
wave amplitude, $\delta c(\underline q)$, must be proportional to the 
longitudinal deformation $\delta\gamma$ (as follows from 
eqn~\ref{mqbal},~\ref{gammadot}):~\footnote{For simplicity we assume 
here that both $c(\underline q)$ and $\delta c(\underline q)$ are real.} 
\begin{equation}\delta c(\underline q)/c_0=-\delta\gamma (\underline q)\label{dcc}\end{equation}
and that the free energy increment, $\delta {\cal F}$, associated with an 
infinitesimal strain $\delta\gamma$ is proportional to the longitudinal stress 
$\sigma_{11}(\underline q)$~\cite{HansDon}:  
\begin{equation}\delta {\cal F}=V\sigma_{11}(\underline q)\delta\gamma (\underline 
q)^{*}\label{dFV}\end{equation}
(note that the NRC are used here as before).  Recalling also that 
$\delta {\cal F}=V\kappa (q)c(\underline q)\delta c(\underline q)^{*}$, as follows from $
{\cal F}\simeq\frac {\kappa V}2\left|c(\underline q)\right|^2$, and using 
eqs.~\ref{dcc},~\ref{dFV}, we find $\sigma_{11}(\underline q)=-\kappa (q)c_0c(\underline 
q)$.  An increment 
of the longitudinal stress due to an increment of $c(\underline q)$ therefore is 
$\delta\sigma_{11}(\underline q)=-\kappa (q)c_0\delta c(\underline q)$, which gives on using again eqn~\ref{dcc}:  
\begin{equation}\delta\sigma_{11}(\underline q)=\kappa (q)c_0^2\delta\gamma (\underline 
q)\label{dsig}\end{equation}
By definition (cf eqn.~\ref{e0}, see also eqn~\ref{Ledef}) the prefactor 
before $\delta\gamma$ in the above equation must be equal to the genuinely static 
longitudinal modulus $L_e(q)$, hence 
\begin{equation}L_e(q)=\kappa (q)c_0^2=\frac {c_0T}{S_2(q)}\label{Legen}\end{equation}
in agreement with the last equation of the Appendix B.  Note that 
eqn~\ref{Legen} generalizes the classical compressibility equation (cf 
eqn~\ref{IIe4}) to systems with mass polydispersity for which the 
function $S_2(q)$ is different from the classical structure factor $S(q)$ 
(cp eqn~\ref{eSq} and~\ref{Sqpoly}).  

It is also worth noting that the generalized compressibility equation 
was derived in Appendix B based on a simple result for the 
cross-correlations between the longitudinal stress and concentration 
fluctuations (eqn~\ref{ee6}).  Below we provide an alternative 
derivation of eqn~\ref{ee6} which will be useful for the discussion on 
the transverse modulus $M_e(q)$ in section~\ref{sec34}.  

\subsection{Cross-correlations between the longitudinal stress and 
concentration fluctuations}

To start with let us recall the microscopic definition of the 
$\underline q$-dependent stress valid for a system of spherical particles with 
pairwise interactions (cf ref.~\cite{BalZop} and eqn  (70) of 
ref.~\cite{Polymers2024} or eqn (13) of ref.~\cite{SM2025}): 
\begin{equation}\sigma_{\alpha\beta}(\underline q)=\sigma_{\alpha\beta}^{vir}(\underline 
q)+\sigma_{\alpha\beta}^{id}(\underline q)\label{IIes}\end{equation}
where $\sigma_{\alpha\beta}^{id}(\underline q)$ is the ideal-gas stress due particle momenta (cf the 
last term in eq.~\ref{G2}), and $\sigma_{\alpha\beta}^{vir}(\underline q)$ is the virial part of the stress 
due to pairwise interactions of the particles:  
\begin{equation}\sigma_{\alpha\beta}^{vir}(\underline q)=\frac 1V\sum_{i\neq j}u'_{
ij}(r)\frac {r_{\alpha}r_{\beta}}{{\rm i}\underline q\cdot\underline r}\frac 1re^{
-{\rm i}\underline q\cdot\underline r_i}\label{evir}\end{equation}
(the sum here involves all $ij$ pairs).  The virial stress $\sigma_{\alpha\beta}^{
vir}(\underline q)$ 
corresponds to the term before the last in eq.~\ref{G2}.  

For the longitudinal stress component (with $\alpha =\beta =1$; note that we 
use the NRC so that the axis 1 is parallel to the wavevector $\underline q$) 
eqn~\ref{evir} can be simplified as 
\begin{equation}\sigma_{11}^{vir}(\underline q)=\frac 1{{\rm i}qV}\sum_{i\neq j}
u'_{ij}(r)\frac xre^{-{\rm i}\underline q\cdot\underline r_i}=\frac 1{{\rm i}qV}
\sum_if_{xi}e^{-{\rm i}qx_i}\label{evir2}\end{equation}
where $x=\underline r\cdot\underline q/q$ is the coordinate along axis 1, and $f_{
xi}$ is the 
longitudinal projection (onto the $x$-axis) of the total force acting on 
particle $i$ from all other particles: 
\begin{equation}f_{xi}=-\frac {\partial U(\Gamma )}{\partial x_i}\label{fxi}\end{equation}
Here $U(\Gamma )$ is the total interaction energy of all particles and $\Gamma$, the 
set of all their coordinates.  Taking into account that the equilibrium 
system is macroscopically uniform which implies that a translation 
of each particle by a vector $\underline u_{tr}$ conserves the total interaction 
energy leading to statistically the same ensemble, we find the 
cross-correlation function of the virial stress with concentration (cf 
eqn~\ref{cqt}) using eqn~\ref{evir2}:  
\begin{equation}\left<\sigma_{11}^{vir}(\underline q,t')c^{*}(\underline q,t')\right
>=\frac T{V^2}\sum_{i\neq j}\frac {m_j}{\bar {m}}\left<e^{{\rm i}q\left(x_j-x_i\right
)}\right>=\frac {Tc_0}V\left[S_1(q)-1\right]\label{esigc}\end{equation}
where $\frac {\partial U(\Gamma )}{\partial x_i}\exp\left(-U(\Gamma )/T\right)=-
T\frac {\partial}{\partial x_i}\exp\left(-U(\Gamma )/T\right)$ was used.  The first 
term in square brackets in the very rhs of eqn~\ref{esigc} comes 
from the sum over all $ij$ pairs, while the second term there 
corresponds to the `diagonal' sum, $\sum_{i=j}$, which should be taken out.  
Here the function $S_1(q)$ is the modified structure factor defined as 
(cf eqn~\ref{eSq},~\ref{Sqpoly}) 
\begin{equation}S_1(q)=\frac 1N\sum_{i,j}\frac {m_j}{\bar {m}}\left<e^{{\rm i}q\left
(x_j-x_i\right)}\right>\label{S1q}\end{equation}
Turning instead to the ideal-gas stress, we find:
\begin{equation}\left<\sigma_{11}^{id}(\underline q)c^{*}(\underline q)\right>=-\frac {
Tc_0}VS_1(q)\label{esigic}\end{equation}
(recall that $\sigma_{11}^{id}(\underline q)$ is defined by the last in eqn~\ref{G2}). Summing 
both sides of eqn~\ref{esigc} and~\ref{esigic} we arrive at 
\[\left<\sigma_{11}(\underline q)c^{*}(\underline q)\right>=-c_0T/V\]
which coincides with eqn~\ref{ee6} in Appendix B. 

\subsection{\label{sec34}Relation between the equilibrium transverse modulus and 
the structure factors}

In order to obtain the transverse equilibrium modulus $M_e(q)$ we need 
to calculate the cross-correlation function of the lateral stress, 
$\sigma_{22}(\underline q)$, and concentration, $\left<\sigma_{22}(\underline q)
c^{*}(\underline q)\right>$.  The stress-concentration 
correlation function can be split into two parts, the virial, 
$\left<\sigma_{22}^{vir}(\underline q)c^{*}(\underline q)\right>$, and the ideal, $\left
<\sigma_{22}^{id}(\underline q)c^{*}(\underline q)\right>$.  The latter part can be 
easily obtained since the pre-averaging of $\sigma_{22}^{id}(\underline q)$ over the velocities 
gives exactly the same result as for the longitudinal stress 
component leading to (cf eqn~\ref{esigic}) 
\begin{equation}\left<\sigma_{22}^{id}(\underline q)c^{*}(\underline q)\right>=-\frac {
Tc_0}VS_1(q)\label{esi22c}\end{equation}
Turning to the virial stress part, 
the correlation function 
\begin{equation}C_{vir}(q)\equiv\left<\sigma_{22}^{vir}(\underline q)c^{*}(\underline 
q)\right>\label{Cvir}\end{equation}
does not seem to be related to a conventional structure factor and, 
therefore, should be obtained by computer simulations in analogy with 
the procedure employed to get the standard structure factor $S(q)$.  
Here the numerical simulation task is expected to be somewhat more 
time-consuming since for any particle $i$ one has to take into account 
the contributions of all neighbors interacting with the particle.  It 
may be convenient to define a function related to the correlation 
function $C_{vir}(q)$ in analogy with eqn~\ref{esigic} as 
\begin{equation}\tilde {S}(q)=-\frac V{Tc_0}C_{vir}(q)\label{CvirS}\end{equation}
The function $\tilde {S}(q)$ will be referred to as the lateral structure factor.  
Note that this function is independent of $V$ in the thermodynamic 
limit just like the standard structure factor $S(q)$.  The `minus' sign 
in eqn~\ref{CvirS} serves to render $\tilde {S}(q)$ positive since (at least at 
low $q$) the longitudinal and lateral stress components should be close 
to each other, and both stresses should anticorrelate with 
concentration (as a higher concentration in liquids must result in a 
higher pressure, hence lower stress).  Combining eqn~\ref{esi22c} 
and~\ref{CvirS} we get 
\begin{equation}\left<\sigma_{22}(\underline q)c^{*}(\underline q)\right>=-\frac {
Tc_0}V\left[S_1(q)+\tilde S(q)\right]\label{es22c}\end{equation}
Finally, using the argument presented at the end of Appendix B below 
eqn~\ref{ee6}, we find that the correlator $\left<AB^{*}\right>$ (with $A=\sigma_{
22}(\underline q)$ and 
$B=c(\underline q)$, see eqn~\ref{lam},~\ref{cqcq}) is given by the rhs of 
eqn~\ref{es22c}.  Taking also into account that $\lambda =-M_e(q)/c_0$ (instead 
of eqn~\ref{lam2}) we thus obtain 
\begin{equation}M_e(q)=c_0T\frac {S_1(q)+\tilde {S}(q)}{S_2(q)}=L_e(q)\left[S_1(
q)+\tilde S(q)\right]\label{eMe}\end{equation}
where eqn~\ref{Legen} was used in the last step.  This shows that 
$M_e(q)$ and $L_e(q)$ generally exhibit different dependence on $q$ and other 
parameters, except at $q\to 0$ where $M_e(q)-L_e(q)\to 0$ implying that 
$S_1(q)+\tilde {S}(q)\to 1$ (cf Sect.  6.4 in ref.~\cite{Polymers2024}).  
Furthermore note that $M_e(q)=L_e(q)+{\cal O}(q^2)$ at small $q$ as follows from 
$G_e(q)=0$ and eqn (121) in ref.~\cite{Polymers2024}.  $ $ 

\subsection{\label{Doem} Discussion on equilibrium moduli}

In the above sections we demonstrated that the genuinely static 
(equilibrium) $q$-dependent elastic moduli (corresponding to infinite 
time shift $t$), namely the longitudinal and transverse moduli, 
$L_e(q)=L(q,t\to\infty )$ and $M_e(q)=M(q,t\to\infty$), can be related to 
stress-concentration cross-correlation functions at zero time shift 
(reflecting {\em structural\/} cross-correlations).  More 
precisely we found that $L_e(q)$ can be obtained in the general case 
based on the static structure factor $S_2(q)$ defined in eqn~\ref{Sqpoly} 
(cf eqn~\ref{Legen}), while $M_e(q)$ can be expressed in terms of three 
correlation functions, the structure factors $S_1(q)$ and $S_2(q)$ (cf 
eqn~\ref{S1q} and~\ref{Sqpoly}), and the lateral structure factor $\tilde {S}(q)$ 
related to cross-correlations of the transverse stress and 
concentration (cf eqn~\ref{CvirS}).  

Thus, we established a link between the long-time moduli and 
structural correlations of concentration (reduced density) and stress.  
Importantly, this link provides an independent alternative way to 
calculate $q$-dependent equilibrium moduli, $L_e(q)$ and $M_e(q)$.  Calculation 
of $q$-dependent relaxation moduli (like $L(q,t)$ and $M(q,t)$) proved to be 
a formidable problem~\cite{JCP164505,Jan2,Jan7}.  This problem gets 
simpler in the limit of vanishing $q$, $q\to 0$, where $L_e(q)$ and $M_e(q)$ 
become equal.~\cite{SoftM7867,Polymers2024} However, even in the 
limiting case, $q\to 0$, calculation of the moduli in the long-time regime 
is a difficult task since (i) very long simulation runs are needed, and 
(ii) the statistics becomes progressively poorer for long time-shifts 
(in particular, close to the glass-transition temperature $T_g$) leading to 
larger error bars.~\cite{MolPhys2881,SoftM7867} The method proposed 
in this paper is free from these difficulties and should allow for a 
rather facile calculation of the equilibrium moduli $L_e(q)$ and $M_e(q)$ 
(based on eqn~\ref{eMe},~\ref{Legen}) using moderately long simulation 
runs like those required to obtain the standard structure factor $S(q)$ 
(cf eqn~\ref{eSq}).  

Furthermore, in the present section we generalized the classical 
equation for $L_e(q)$ (cf eqn~\ref{IIe4}) to systems with mass 
dispersity.  The obtained results are valid for $q\neq 0$ including the 
regime $q\to 0$, where both the structure factors and the equilibrium 
moduli show well-defined limits as argued in ref.~\cite{Polymers2024}.  
Noteworthily, eqn~\ref{eMe} derived above allows to calculate $M_e(q)$ 
in the general case (including polydisperse systems with different 
types of particle dispersity).  We also established that both moduli, 
$L_e(q)$ and $M_e(q)$, depend not only on the interaction potentials between 
the particles but also on their masses.  A natural question then 
arises:  why the static (equilibrium) moduli at a finite $q$ depend on 
particle masses.  The reason is that a static longitudinal deformation 
is proportional to the concentration wave amplitude (cf eqn~\ref{dcc}), 
whose definition involves the mass factors (cf eqn~\ref{cqt}).  This 
point is further clarified in parts 1 and 2 of the discussion 
section~\ref{DS}.  
   
\section{\label{sIII} The general relation between 
the elasticity tensor and the stress correlation functions}

As mentioned in the Introduction, in this section we present a new 
derivation of the relation, eq.~\ref{e2}, between the stress correlation 
tensor, $C_{\alpha\beta\gamma\delta}(\underline q,t)$, and the elasticity tensor, $
E_{\alpha\beta\gamma\delta}\left(\underline q,t-t'\right)$ (cf 
eqs.~\ref{e7},~\ref{f5}) which does not involve any consideration of non-steady 
dynamics employed in ref.~\cite{Polymers2024}.  In particular, it is 
shown below that eqn~\ref{e2} indeed follows from eqn~\ref{f5} 
linking the generalized relaxation moduli with correlations of stress 
noise.  

\subsection{~\label{TheRel}The relation for longitudinal components}

Let us consider an amorphous system (a supercooled liquid) with the 
classical dynamics.  The total stress field at wavevector $\underline q$, $\sigma_{
\alpha\beta}(\underline q,t)$, 
defined in eq.~\ref{G2}, is always the sum of the time-dependent 
deterministic stress, 
$ $$\sigma_{\alpha\beta}^d(\underline q,t)$, and the stress noise, $\sigma_{\alpha
\beta}^n(\underline q,t)$, cf eqn~\ref{e6}.  
Using the momentum equation~\ref{e11} and 
eqn~\ref{e7} we get the following relation between $\sigma^d$ and $\sigma$ in terms 
of their Fourier transforms (FTs) with respect to time indicated 
solely by the argument, frequency $\omega$ instead of $t$:  
\begin{equation}\sigma^d_{\alpha\beta}(\underline q,\omega )=\frac {{\rm i}}{\rho_
0\omega}E_{\alpha\beta\gamma\delta}\left(\underline q,\omega\right)\sigma_{\gamma
\epsilon}(\underline q,\omega )q_{\epsilon}q_{\delta}\label{e12a}\end{equation}
where $\alpha ,\beta ,\gamma ,..$ denote Cartesian components, $\underline q$ is the wavevector and 
$\rho_0=\bar {m}c_0$ is the mean mass density.  Here the FT of a $t$-dependent 
function $f(t)$ is defined as:  
\begin{equation}f(\omega )=\int_{-\infty}^{\infty}f(t)\exp(-{\rm i}\omega t)e^{-
\epsilon\left|t\right|}{\rm d}t\label{e10}\end{equation}
where $\epsilon =0^{+}$ is an infinitesimal positive number needed to impart 
convergence.~\footnote{ More precisely, $\Delta t=1/\epsilon$ can be treated as the 
sampling time that is assumed to be very long, much longer than the 
terminal relaxation time $\tau_{\alpha}$ of the system, or formally $\Delta t\to
\infty$ and 
$\epsilon\to 0$.  More generally, the factor $e^{-\epsilon\left|t\right|}$ in eqn~\ref{e10} can be 
replaced by any real weight function $w_{\epsilon}(t)$ such that $w_{\epsilon}(t
)\to 1$ at $\epsilon\to 0$ 
and $w_{\epsilon}(t)\to 0$ at $\left|t\right|\to\infty$; in this case $\Delta t=
\int_{-\infty}^{\infty}w_{\epsilon}(t)^2{\rm d}t$.} 
  
Eq.~\ref{e12a} can be rewritten as~\footnote{Note that both $\sigma_{\gamma\epsilon}
(\underline q,t)$ 
and $\sigma^d_{\alpha\beta}(\underline q,t)$ can show high-frequency fluctuations.} 
\begin{equation}\sigma^d_{\alpha\beta}(\underline q,\omega )=\frac 1{\rho_0\omega^
2}\breve {E}_{\alpha\beta\gamma\delta}\left(\underline q,s={\rm i}\omega\right)\sigma_{
\gamma\epsilon}(\underline q,\omega )q_{\epsilon}q_{\delta}\label{e12}\end{equation}
where the `breve' sign indicates the Laplace-Carson 
transform~\cite{RR} (known also as the 
`$s$-transform'~\cite{Polymers2024,JCP164505,PRE2023}) of a 
time-dependent function:  
\begin{equation}\breve {E}_{\alpha\beta\gamma\delta}\left(\underline q,s\right)=
s\int_0^{\infty}E_{\alpha\beta\gamma\delta}\left(\underline q,t\right)e^{-\epsilon 
t}e^{-st}{\rm d}t\label{e120}\end{equation}
and $\epsilon =0^{+}$.  In order to show that eqn~\ref{e2} follows from 
eqn~\ref{f5} it is instructive to start with a particular (simple but 
important) case when the wavevector $\underline q$ is oriented along axis 1, and 
we are interested only in longitudinal components of all tensors, like 
$L(q,t)\equiv E_{1111}(\underline q,t)$ and $C_L(q,t)\equiv C_{1111}(\underline 
q,t)$, cf 
refs~\cite{Polymers2024,SM2018}.  Using eqn~\ref{e6},~\ref{e12} we get 
\begin{equation}\sigma^n_{11}(\underline q,\omega )=\left[1+\varkappa\breve L(q,
s={\rm i}\omega )\right]\sigma_{11}(\underline q,\omega ),\mbox{\rm \ where }\varkappa
\equiv q^2/\left(\rho_0s^2\right)\label{e13}\end{equation}
and the function $\breve {L}(q,s)$ is the Laplace-Carson transform (cf 
eq.~\ref{e120}) of the time-dependent longitudinal relaxation modulus 
$L(q,t)$.  

In view of eqn~\ref{e4},~\ref{e3}, the Fourier transforms of $C$ and 
$C^n$ are (due to the Wiener-Khinchin theorem)
\begin{equation}C_{\alpha\beta\alpha'\beta'}(\underline q,\omega )\simeq\frac V{
T\Delta t}\left<\sigma_{\alpha\beta}(\underline q,\omega )\sigma_{\alpha'\beta'}
(\underline q,\omega )^{*}\right>\label{e14}\end{equation}
\begin{equation}C_{\alpha\beta\alpha'\beta'}^n(\underline q,\omega )\simeq\frac 
V{T\Delta t}\left<\sigma_{\alpha\beta}^n(\underline q,\omega )\sigma_{\alpha'\beta'}^
n(\underline q,\omega )^{*}\right>\label{e15}\end{equation}
where star $(^{*}$) means complex conjugate (note that 
$\sigma_{\alpha\beta}(\underline q,\omega )^{*}=\sigma_{\alpha\beta}(-\underline 
q,-\omega )$), $\Delta t=1/\epsilon$ (cf.  eqn~\ref{e10}), and the above 
two equations become asymptotically exact for $\epsilon\to 0$~since both 
correlation functions, $C$ and $C^n$, decay to 0 at $\left|t\right|\gg\tau_{\alpha}$, where $
\tau_{\alpha}$ is 
the terminal $(\alpha$-relaxation) time.  

On using the last 3 equations we obtain:  
\begin{equation}C_L^n(q,\omega )\equiv C_{1111}^n(\underline q,\omega )=\left[1+
\varkappa\breve L(q,s)\right]\left[1+\varkappa\breve L(q,-s)\right]C_L(q,\omega 
)\label{e16}\end{equation}
where $s={\rm i}\omega$ (and $\omega\neq 0$ is real) here and below (until the end of 
section~\ref{gencase}).  Furthermore, on using also 
eqn~\ref{e10},~\ref{e120} we find 
\begin{equation}{\rm i}\omega C_L(q,\omega )=\breve {C}_L(q,s)-\breve {C}_L(q,-s
)\label{e16a}\end{equation}
while using in addition eqn~\ref{f5} we obtain
\begin{equation}{\rm i}\omega C_L^n(q,\omega )=\breve {L}(q,s)-\breve {L}(q,-s)\label{e16b}\end{equation}
The above two relations allow to transform eqn~\ref{e16} as 
\begin{equation}\breve {L}(q,s)-\breve {L}(q,-s)=\left[1+\varkappa\breve L(q,s)\right
]\left[1+\varkappa\breve L(q,-s)\right]\left(\breve C_L(q,s)-\breve C_L(q,-s)\right
)\label{e17}\end{equation}
The latter equation is always satisfied if 
\begin{equation}\breve {L}(q,s)=\left[1+\varkappa\breve L(q,s)\right]\breve {C}_
L(q,s)\label{e18}\end{equation}
Note that the above equation, which is valid for $s={\rm i}\omega$, defines $C_L$ 
for a given $L$.  Importantly, there are no other solutions to 
eqn~\ref{e17} since in view of eqn~\ref{e16a} (and for a given $L$) it 
uniquely defines the Fourier transform $C_L(q,\omega )$ and therefore also 
both $C_L(q,t)$ and $\breve {C}_L(q,s)$.~\footnote{This statement is rigorous if the 
factors in square brackets (in eqn~\ref{e17}) are non-zero for any 
$s={\rm i}\omega$ with real $\omega$.  The latter condition is indeed satisfied since otherwise the 
lhs of eqn~\ref{e17}, which is proportional to the loss modulus 
$L^{\prime\prime}(q,\omega )\equiv\omega\int_0^{\infty}L(q,t)\cos(\omega t){\rm d}
t$, would vanish, which is impossible since 
$L^{\prime\prime}$ must be positive (this point is further discussed in 
section~\ref{gencase}).  Therefore, $1+\varkappa\breve {L}(q,s)\neq 0$ which also means 
that eqn~\ref{e18} defines a unique $\breve {C}_L(q,s)$ in terms of $\breve {L}(
q,s)$.} 
Finally note that eqn~\ref{e18} is in harmony with eqn~\ref{e2} for 
the longitudinal component of the stress autocorrelation function, 
$\breve {C}_{1111}(\underline q,s)=\breve {C}_L(q,s)$.  Therefore we demonstrated that eqn~\ref{e2} for 
the longitudinal components of $\breve {E}$ and $\breve {C}$ does follow from eqn~\ref{f5}.  
This result supports the general eqn~\ref{e2}.

\subsection{\label{gencase} The general case}

Let us turn to the general case of the relation between the 
tensorial memory function $E_{\alpha\beta\alpha'\beta'}(\underline q,t)$ and the tensor of stress 
correlation functions $C_{\alpha\beta\alpha'\beta'}(\underline q,t)$.  On using eqn~\ref{e12},~\ref{e6} 
we get (for real $\omega\neq 0$ and $s={\rm i}\omega$)
\begin{equation}\sigma^n_{\alpha\beta}(\underline q,\omega )={\cal A}_{\alpha\beta
\gamma\epsilon}(\underline q,s)\sigma_{\gamma\epsilon}(\underline q,\omega )\label{f8}\end{equation}
where 
\begin{equation}{\cal A}_{\alpha\beta\gamma\epsilon}(\underline q,s)\equiv\delta_{
\alpha\gamma}\delta_{\beta\epsilon}+\varkappa\breve {E}_{\alpha\beta\gamma\delta}\left
(\underline q,s\right)\hat {q}_{\delta}\hat {q}_{\epsilon}\label{f9}\end{equation}
and $\hat{\underline q}=\underline q/q$.  Then, taking into account eqn~\ref{e14},~\ref{e15}, we 
obtain a relation between the correlation functions of the full stress 
and the stress noise:  
\begin{equation}C^n_{\alpha\beta\alpha'\beta'}(\underline q,\omega )={\cal A}_{\alpha
\beta\gamma\epsilon}(\underline q,s){\cal A}_{\alpha'\beta'\gamma'\epsilon'}(\underline 
q,-s)C_{\gamma\epsilon\gamma'\epsilon'}(\underline q,\omega )\label{f10}\end{equation}
(Note that $\breve {E}_{\alpha\beta\gamma\delta}\left(\underline q,-s\right)=\breve {
E}^{*}_{\alpha\beta\gamma\delta}\left(\underline q,s\right)$ for $s={\rm i}\omega$ according to 
eq.~\ref{e120}, and hence ${\cal A}$$_{\alpha\beta\gamma\delta}\left(\underline q,
-s\right)={\cal A}^{*}_{\alpha\beta\gamma\delta}\left(\underline q,s\right)$.)  Furthermore, 
using eqn~\ref{f5},~\ref{e8} we find (again for $s={\rm i}\omega$) 
\begin{equation}sC_{\alpha\beta\alpha'\beta'}^n(\underline q,\omega )=\breve {E}_{
\alpha\beta\alpha'\beta'}(\underline q,s)-\breve {E}_{\alpha\beta\alpha'\beta'}(\underline 
q,-s)\label{f12}\end{equation}

Eqn~\ref{f10} can be considered as a linear matrix equation 
\begin{equation}C^n_{\alpha\beta\alpha'\beta'}(\underline q,\omega )={\cal M}_{\alpha
\beta\alpha'\beta'\gamma\epsilon\gamma'\epsilon'}C_{\gamma\epsilon\gamma'\epsilon'}
(\underline q,\omega )\label{f14}\end{equation}
where ${\cal M}={\cal A}{\cal A}^{*}$ is given by the product of two ${\cal A}$-tensors in the rhs of 
eqn~\ref{f10} defining $C^n$ in terms of $C$.  For a given $C^n$ 
eqn~\ref{f14} can be solved for $C$ using the standard linear algebra, 
so that eventually (also taking into account eqn~\ref{f12}) the stress 
correlation tensor $C$ can be expressed in terms of the generalized 
relaxation moduli, $E$.  More precisely, this is true if the solution of 
eqn~\ref{f14} for $C$ is unique which means that the matrix ${\cal M}$ is not 
degenerate.  In other words, for any nonzero 4th-rank tensor $C$, ${\cal M}\cdot 
C$ 
is also nonzero.  To prove this statement let us assume the opposite:  
that there exists such $C\neq 0$ that ${\cal M}\cdot C=0$.  The latter assumption 
implies that either $X\equiv {\cal A}^{*}\cdot C=0$ or, else, $X\neq 0$ but ${\cal A}
\cdot X=0$.  Let's 
consider the second option, ${\cal A}\cdot X=0$ (the reasoning for ${\cal A}^{*}
\cdot C=0$ is 
analogous):  
\begin{equation}{\cal A}_{\alpha\beta\gamma\epsilon}(\underline q,s)X_{\alpha'\beta'
\gamma\epsilon}=0\label{f15}\end{equation}
As $X\neq 0$ there exist such $\alpha',\beta'$ that $Y_{\gamma\epsilon}\equiv X_{
\alpha'\beta'\gamma\epsilon}$ is non-zero, while 
${\cal A}\cdot Y=0$.  Then, imposing that 
$\sigma_{\gamma\epsilon}(\underline q,\omega )={\rm c}{\rm o}{\rm n}{\rm s}{\rm t}\,
Y_{\gamma\epsilon}$ in eqn~\ref{f8} we find 
that $\sigma^n_{\alpha\beta}(\underline q,\omega )=0$.  Therefore we arrive at the dynamical state with 
a stress fluctuation wave (with frequency $\omega$ and wavevector $\underline q$) 
which stays forever without decay since eqn~\ref{f8} in this case 
says that the mechanical balance is maintained by the deterministic 
stress only.  It means that there is no dissipation in this dynamical 
state, which is impossible according to the basic principle of 
thermodynamics stating that the dissipation rate ${\cal D}$ must be positive 
in a non-equilibrium system (like the system we consider which 
shows permanent oscillations), hence the entropy of any such 
thermally isolated system must increase in time (according to the 
second law of thermodynamics).  Therefore $\sigma^n_{\alpha\beta}(\underline q,\omega 
)=0$ is not 
allowed (for any $q\neq 0$, $\omega\neq 0$).  

This conclusion is also in line with the fact that a noise, the stress 
noise in the present case, is intimately related to the dissipation and 
friction (cf Appendix A).  Noteworthily, eqn~\ref{af10} proves that 
the energy dissipation rate ${\cal D}$ cannot be negative (in a system 
perturbed by an external force).  To sum up, the above physical 
argument ensures that the matrix ${\cal M}$ is not degenerate and so 
eqn~\ref{f14} defines a unique $C$ in terms of $C^n$ and $E$.  

Let us now proceed with the derivation of eqn~\ref{e2}.  
Taking into account that the Fourier transform of the full stress 
correlation function is (cf. eq.~\ref{e16a})
\[C_{\alpha\beta\gamma\delta}(\underline q,\omega )=\frac 1s\left[\breve C_{\alpha
\beta\gamma\delta}(\underline q,s)-\breve C_{\alpha\beta\gamma\delta}(\underline 
q,-s)\right]\]
with $s={\rm i}\omega$ and real $\omega\neq 0$ as before, and using eqn~\ref{f12} we rewrite 
eqn~\ref{f10} as
\begin{equation}\breve {E}_{\alpha\beta\alpha'\beta'}(\underline q,s)-\breve {E}_{
\alpha\beta\alpha'\beta'}(\underline q,-s)={\cal A}_{\alpha\beta\gamma\epsilon}(\underline 
q,s){\cal A}_{\alpha'\beta'\gamma'\epsilon'}(\underline q,-s)\left[\breve C_{\gamma
\epsilon\gamma'\epsilon'}(\underline q,s)-\breve C_{\gamma\epsilon\gamma'\epsilon'}
(\underline q,-s)\right]\label{f18}\end{equation}
Replacing the ${\cal A}$-factors in the above equation with expressions in the 
rhs of eqn~\ref{f9} and omitting (by brutal force) in the resultant 
equation all terms with the second argument `$-s$' we 
get:~ 
\begin{equation}\breve {E}_{\alpha\beta\alpha'\beta'}(\underline q,s)={\cal A}_{
\alpha\beta\gamma\epsilon}(\underline q,s)\breve {C}_{\gamma\epsilon\alpha'\beta'}
(\underline q,s)\label{f19}\end{equation}
and an equivalent equation can be obtained by omitting all terms with 
argument $s$ in eqn~\ref{f18}.\footnote{{Note that eq.~\ref{f19} is a
{\bf guess} equation.  As explained below, eqn~\ref{f18} does follow from 
eq.~\ref{f19}.  Therefore, given that both eqn~\ref{f18} and 
eqn~\ref{f19} have a unique solution, they must be equivalent.}} 
Furthermore, one can easily show (using the major symmetries of $C$ 
and $E$ tensors) that if eqn~\ref{f19} is valid, the basic eqn~\ref{f18} is 
satisfied.  Taking also into account that (for the reasons presented 
below eqn~\ref{f14}) eqn~\ref{f18} has a unique solution (defining the 
function $C$ once the elasticity tensor function $E$ is given), and the 
same is true for eqn~\ref{f19} (for the similar reasons) we deduce 
that the unique solution of the basic eqn~\ref{f18} derived above must 
also satisfy eqn~\ref{f19} which can be written in a more explicit 
form as:  
\begin{equation}\breve {C}_{\alpha\beta\alpha'\beta'}(\underline q,s)=\breve {E}_{
\alpha\beta\alpha'\beta'}(\underline q,s)-\frac {q_{\delta}q_{\epsilon}}{\rho_0s^
2}\breve {E}_{\alpha\beta\gamma\delta}(\underline q,s)\breve {C}_{\gamma\epsilon
\alpha'\beta'}(\underline q,s)\label{e2}\end{equation}
where the `breve' sign indicates the Laplace-Carson transform.  

To conclude, we thus proved, without invoking any constraints on the 
flow field (imposed in refs.~\cite{Polymers2024,SM2025}), that 
eqn~\ref{e2} indeed follows from eqn~\ref{f5} which was discussed in 
section~\ref{ECn}.  Equations~\ref{e2} and~\ref{f5} are therefore 
generally valid for all equilibrated amorphous systems with 
time-translational invariant (and unconstrained) Newtonian dynamics 
(as was anticipated in refs.~\cite{SM2018,Polymers2024,SM2025}).  It is 
also worth noting that the above derivation is applicable also to 
polydisperse systems (in particular those with mass or size 
polydispersity of particles) and to systems with arbitrary thermal 
conductivity (including isothermic and adiabatic systems as the 
limiting cases).  

\section{\label{sI} The stress noise vs.  the reduced deviatoric stress
of the projected dynamics: do they coincide} 

\subsection{\label{sI1}General relations for the stress noise and the 
reduced deviatoric stress} 

As shown above (in section~\ref{sIII}) the correlation function 
$C^n_{\alpha\beta\gamma\delta}(\underline q,t)$ of the stress noise, $\sigma^n_{
\alpha\beta}(\underline q,t)$, is related to the generalized 
elasticity tensor (cf eqn~\ref{e3},~\ref{f5}).  Importantly, the $C^n$ 
correlation tensor was recently identified~\cite{SM2025} with the 
memory kernel $M_{\alpha\beta\gamma\delta}(\underline q,t)$~\cite{FuchsJCP2018,FuchsPRL2017} of the 
Zwanzig-Mori (ZM) projection operator 
formalism~\cite{HansDon,Goetze,BalZop,Zwan,Mori}, cf eqn~\ref{MCn}, 
defined in terms of the reduced deviatoric stress 
$\sigma^{rd}_{\alpha\beta}(\underline q,t)$:~\footnote{See eqn (56) of ref.~\cite{SM2025} and 
refs.~\cite{FuchsPRL2017,FuchsJCP2018}.} 
\begin{equation}M_{\alpha\beta\gamma\delta}(\underline q,t)=\frac VT\left<\sigma_{
\alpha\beta}^{rd}(\underline q,t)\sigma_{\gamma\delta}^{rd}(\underline q,0)^{*}\right
>\label{Mdef}\end{equation}
where 
\begin{equation}\sigma_{\alpha\beta}^{rd}(\underline q,t)\equiv {\cal R}'(t)\sigma_{
\alpha\beta}(\underline q)\label{Ie4a}\end{equation}
is the reduced deviatoric stress, $\sigma_{\alpha\beta}(\underline q)$ is the stress tensor 
implicitly considered as a function of a microstate,~\footnote{Note 
that all variables (like $\sigma_{\alpha\beta}(\underline q)$) in the ZM formalism are considered as 
functions of a microstate $\Gamma$ in the Hamiltonian phase space.} and 
\begin{equation}{\cal R}'(t)={\cal Q}e^{{\rm i}{\cal L}{\cal Q}t},~{\cal Q}=1-{\cal P}\label{Rprime}\end{equation}
is the reduced time-evolution operator (cf eqn (52) in~\cite{SM2025}).  
Here ${\cal L}$ is the classical Liouvillean~\cite{Goetze} and ${\cal Q}$ is the 
complementary operator to the projection operator ${\cal P}$ which projects 
onto the conserved variables, viz.  mass density and momentum 
density (cf eqn~\ref{Jqt}; see also eqn (49) in ref.~\cite{SM2025}).    

In view of the definitions, eqn~\ref{e3},~\ref{Mdef}, eqn~\ref{MCn} 
hints at a close relationship between the stress noise $\sigma^n(q,t)$ and 
$\sigma^{rd}(q,t)$.  Indeed, it was shown~\cite{SM2025} that if the constrained 
dynamics is imposed at $t<0$ (ie the flow at wavevector $\underline q$ is 
completely suppressed due to external force, eq.~\ref{eqF}, applied at 
$t<0$ and leading to $\underline J(\underline q,t)=0$), but then the constraint is completely 
released at $t\geq 0$, the two stress fluctuation functions, $\sigma^n_{\alpha\beta}
(\underline q,t)$ and 
$\sigma^{rd}_{\alpha\beta}(\underline q,t)$, must be equal to each other at $t>0$.  Below we show, 
however, that such an equivalence does not hold any more for the 
classical Newtonian (unconstrained) dynamics, thus shedding a new 
light on the issue concerning the stationarity of the two processes.  

As we discussed in section~\ref{sIII}, the total stress tensor $\sigma_{\alpha\beta}
(\underline q,t)$ 
can be represented (cf eqn~\ref{e6}) as a sum of the deterministic 
stress, $\sigma_{\alpha\beta}^d(\underline q,t)$ (related to the flow velocity field, cf eqn~\ref{e7}) 
and the stress noise, $\sigma_{\alpha\beta}^n(q,t)$.  On 
the other hand, within the ZM approach the total stress can be split 
in a different way 
\begin{equation}\sigma_{\alpha\beta}(\underline q,t)=\sigma_{\alpha\beta}^{rd}(\underline 
q,t)+\sigma_{\alpha\beta}^p(\underline q,t)+\int_0^tM_{\alpha\beta\gamma\delta}(\underline 
q,t-t'){\rm i}v_{\gamma}(\underline q,t')q_{\delta}{\rm d}t'\label{Ie4}\end{equation}
(as directly stems from eqn 75, 76 in ref.~\cite{SM2025}).  Note that 
$v_{\gamma}(\underline q,t')$ in the above equation is the microscopic velocity field which 
is not coarse-grained either in space or in time.  Its time-dependence 
reflects thermal fluctuations along a system's trajectory in the phase 
space.  
 
The second term in the rhs of eqn~\ref{Ie4} is the ${\cal P}$-projected stress 
\begin{equation}\sigma_{\alpha\beta}^p(\underline q,t)={\cal R}(t){\cal P}\sigma_{
\alpha\beta}(\underline q)\label{Ie5}\end{equation}
where ${\cal R}(t)$ is the classical time-evolution operator, 
${\cal R}(t)=e^{{\rm i}{\cal L}t}$~\cite{Goetze}.  As follows from eq.~\ref{Ie5} and the 
definition of the projection operator ${\cal P}$, the stress $\sigma_{\alpha\beta}^
p(\underline q,t)$ is simply 
proportional to the density fluctuation:~\footnote{Note that $\rho (\underline q
,t)$ is 
proportional to the concentration fluctuation $c(\underline q,t)$ due to its 
definition, eqn~\ref{cqt}, adopted in the present paper.},~\footnote{In 
agreement with eqn (76) in ref.~\cite{SM2025}.} 
\begin{equation}\sigma_{\alpha\beta}^p(\underline q,t)=-\left[M_e(q)\delta_{\alpha
\beta}+\left(L_e(q)-M_e(q)\right)\hat q_{\alpha}\hat q_{\beta}\right]\rho (\underline 
q,t)/\rho_0\label{Ie6}\end{equation}

Returning to eqn~\ref{Ie4} we observe that the last term in this 
equation depends on the memory kernel defined in 
eqn~\ref{Mdef}.~\footnote{Note that the last term in eqn~\ref{Ie4} 
can be considered as a part of the deterministic stress, eqn~\ref{e7}.} 

Since ${\cal P}+{\cal Q}=1$, another useful splitting of the total stress reads 
\begin{equation}\sigma_{\alpha\beta}(\underline q,t)=\sigma^p_{\alpha\beta}(\underline 
q,t)+\sigma^{dev}_{\alpha\beta}(\underline q,t)\label{spdev}\end{equation}
where the `plain' deviatoric stress is defined as 
\begin{equation}\sigma^{dev}_{\alpha\beta}(\underline q,t)={\cal R}(t){\cal Q}\sigma_{
\alpha\beta}(\underline q)\label{sdev}\end{equation}
Note that in the regime of small wavevector $\underline q$, the projected stress 
$\sigma^p_{\alpha\beta}(\underline q,t)$ tends to the opposite of the isotropic pressure tensor, 
\begin{equation}\sigma^p_{\alpha\beta}(\underline q,t)\simeq -P(\underline q,t)\delta_{
\alpha\beta}\mbox{\rm \ at}~~q\to 0\label{sisoP}\end{equation}
where
\begin{equation}P(\underline q,t)=L_e(q)\rho (\underline q,t)/\rho_0\label{isoP}\end{equation}
for $q\to 0$ represents an isotropic pressure fluctuation.  Eq.~\ref{sisoP} 
comes from eq.~\ref{Ie6} by taking into account that $M_e(q)\to L_e(q)$ in 
the long-wave limit, $q\to 0$ (cf eqn (121) in~\cite{Polymers2024} and 
eq.~\ref{GLM} in section~\ref{sII} above).  Accordingly, in the low $q$ 
regime the plain deviatoric part, $\sigma^{dev}_{\alpha\beta}(\underline q,t)$, can be interpreted as the 
total stress tensor devoid of the isotropic pressure contribution 
(related to density).  
 
As we argued in ref.~\cite{SM2025}, the density-related stress $\sigma^p$ is 
generally not equal to the deterministic stress $\sigma^d$, and consequently, 
the stress noise $\sigma^n$ is different from the plain deviatoric part, $\sigma^{
dev}$.  
By contrast, as mentioned above, the auto-correlation functions of $\sigma^n$ 
and of $\sigma^{rd}$ are equal to each other.~\cite{SM2025} Below we show, 
however, that in the case of the classical dynamics 
(unconstrained at all times), the two stress processes are generally 
different, $\sigma^n(\underline q,t)\neq\sigma^{rd}(\underline q,t)$.  

To prove the latter statement it is enough to demonstrate that the 
stress $\sigma^d$ (which is complementary to $\sigma^n$) is not equal to the sum of 
the last two terms in eqn~\ref{Ie4}, which are complementary to $\sigma^{rd}$.  
Indeed, as follows from eqn~~\ref{e7},~\ref{Ie4},~\ref{Ie5},~\ref{Ie6} 
(cf also eqn~(70) of ref.~\cite{SM2025}), the difference 
$\sigma^{\Delta}_{\alpha\beta}(\underline q,t)\equiv\sigma^{rd}(\underline q,t)-
\sigma^n(\underline q,t)$ is
\begin{equation}\sigma^{\Delta}_{\alpha\beta}(\underline q,t)=\sigma^d_{\alpha\beta}
(\underline q,t)-\left(\sigma_{\alpha\beta}^p(\underline q,t)+\int_0^tM_{\alpha\beta
\gamma\delta}(\underline q,t-t'){\rm i}v_{\gamma}(\underline q,t')q_{\delta}{\rm d}
t'\right)\label{new7}\end{equation}
The above equation also follows from eqn (75), (76) of 
ref.~\cite{SM2025}.  On using eq.~\ref{e7} defining the 
deterministic stress $\sigma^d_{\alpha\beta}(\underline q,t)$, and eqs.  (42), (68), (70) of 
ref.~\cite{SM2025} one gets 
\begin{equation}\sigma^d_{\alpha\beta}(\underline q,t)=\int_{-\infty}^tM_{\alpha
\beta\gamma\delta}(\underline q,t-t'){\rm i}v_{\gamma}(\underline q,t')q_{\delta}
{\rm d}t'+\sigma_{\alpha\beta}^p(\underline q,t)\label{new8}\end{equation}
The above two equations finally lead to
\begin{equation}\sigma^{\Delta}_{\alpha\beta}(\underline q,t)=\int_{-\infty}^0M_{
\alpha\beta\gamma\delta}(\underline q,t-t'){\rm i}v_{\gamma}(\underline q,t')q_{
\delta}{\rm d}t'\label{Ie14}\end{equation}
which is likely to be generally nonzero.  As a result the general 
relation between the stress noise and the reduced deviatoric stress 
reads:  
\begin{equation}\sigma_{\alpha\beta}^n(\underline q,t)=\sigma_{\alpha\beta}^{rd}
(\underline q,t)-\sigma^{\Delta}_{\alpha\beta}(\underline q,t)\label{rdn}\end{equation}
The autocorrelations of $\sigma^{\Delta}$ are considered in the next section.  

\subsection{\label{sI2}Variance of the difference between the reduced deviatoric 
stress and the stress noise}

To quantify the difference between $\sigma^{rd}(\underline q,t)$ and $\sigma^n(\underline 
q,t)$ 
we calculate below its variance 
\begin{equation}C^{\Delta\Delta}_{\alpha\beta\gamma\delta}\equiv\frac VT\left<\sigma^{
\Delta}_{\alpha\beta}(\underline q,t)\sigma^{\Delta}_{\alpha\beta}(\underline q,
t)^{*}\right>_{t=0}\label{Ie16}\end{equation}
which can be compared to the time-independent variances 
of $\sigma^{rd}$ and $\sigma^n$: 
\begin{equation}\frac VT\left<\sigma_{\alpha\beta}^n(\underline q,0)\sigma_{\gamma
\delta}^n(\underline q,0)^{*}\right>=\frac VT\left<\sigma_{\alpha\beta}^{rd}(\underline 
q,0)\sigma_{\gamma\delta}^{rd}(\underline q,0)^{*}\right>=M_{\alpha\beta\gamma\delta}
(\underline q,t=0)\label{Ie17}\end{equation}
Eqn~\ref{Ie14} implies that the variance, 
eqn~\ref{Ie16}, involves the autocorrelation function of the 
$\underline q$-dependent velocity 
\[C^{vv}_{\alpha\beta}(\underline q,t)=\frac VT\left<v_{\alpha}(\underline q,t+t'
)v_{\beta}(\underline q,t')^{*}\right>\]
whose Laplace transform (marked by `tilde' sign) is equal to (cf eqn  
(56) of ref~\cite{Polymers2024}) 
\begin{equation}\tilde {C}^{vv}_{\alpha\alpha'}(\underline q,s)=\delta_{\alpha\beta}
/\left(\rho_0s\right)-\tilde {C}_{\alpha\beta\alpha'\beta'}(\underline q,s)q_{\beta}
q_{\beta'}/\left(\rho_0^2s^2\right)\label{Ie18}\end{equation}
(Note that the Laplace transform of a function $C(t)$ is 
$\tilde {C}(s)\equiv\int_0^{\infty}C(t)e^{-st}{\rm d}t$.)  

The variance, eqn~\ref{Ie16}, can be written using eqn~\ref{Ie14} as
\begin{equation}C^{\Delta\Delta}_{\alpha\beta\gamma\delta}=q_{\beta'}q_{\delta'}
\int_0^{\infty}{\rm d}t'{\rm d}t^{\prime\prime}M_{\alpha\beta\alpha'\beta'}(\underline 
q,t')M_{\gamma\delta\gamma'\delta'}(\underline q,t^{\prime\prime})C^{vv}_{\alpha'
\gamma'}(\underline q,t'-t^{\prime\prime})\label{Ie20}\end{equation}
The above equation can be rewritten in terms of Fourier transforms 
of the relevant functions which are in turn related to the 
corresponding Laplace transforms.  
For example, turning to the last factor in eqn~\ref{Ie20}, which can be 
denoted (omitting the argument $\underline q$) as $C^{vv}_{\alpha'\gamma'}(t)$ with $
t=t'-t^{\prime\prime}$, and recalling that the function $C^{vv}_{\alpha'\gamma'}
(t)$ is 
even, $C^{vv}_{\alpha'\gamma'}(-t)=C^{vv}_{\alpha'\gamma'}(t)$, due to time reversibility~\cite{Polymers2024} we 
represent its Fourier transform as (cf eqn~\ref{e10})
\[C^{vv}_{\alpha'\gamma'}(\omega )=\int_{-\infty}^{\infty}C^{vv}_{\alpha'\gamma'}
(t)e^{-{\rm i}\omega t}e^{-\epsilon\left|t\right|}{\rm d}t=\tilde {C}^{vv}_{\alpha'
\gamma'}(s={\rm i}\omega +\epsilon )+\tilde {C}^{vv}_{\alpha'\gamma'}(s=-{\rm i}
\omega +\epsilon )\]
where $\epsilon =0^{+}$ (as stated below eqn~\ref{e10}) is an infinitesimal 
positive number. 

As a result we get based on the above equations:   
\[C^{\Delta\Delta}_{\alpha\beta\gamma\delta}=q_{\beta'}q_{\delta'}\int_{-\infty}^{
\infty}\frac {{\rm d}\omega}{2\pi}\tilde {M}_{\alpha\beta\alpha'\beta'}(\underline 
q,s={\rm i}\omega +\epsilon )\tilde {M}_{\gamma\delta\gamma'\delta'}(\underline 
q,s=-{\rm i}\omega +\epsilon )\cdot\]
\begin{equation}\left[\tilde C^{vv}_{\alpha'\gamma'}(\underline q,s={\rm i}\omega 
+\epsilon )+\tilde C^{vv}_{\alpha'\gamma'}(\underline q,s=-{\rm i}\omega +\epsilon 
)\right]\label{Ie22}\end{equation}

Let us focus on the longitudinal stress components, $\sigma_{11}(\underline q)$, using the 
NRC with axis 1 parallel to the wavevector $\underline q$.  In this case the only 
relevant component of the $M$-tensor in eqn~\ref{Ie14} is $M_{1111}$ which 
is equal to (cf eqn~\ref{MCn},~\ref{f5},~\ref{GLM}) 
\begin{equation}M_{1111}(\underline q,t)=L(q,t)-L_e(q)\equiv\Delta L(q,t)\label{Ie24}\end{equation}
where $L(q,t)$ is the $(q,t)$-dependent longitudinal modulus and $L_e$ is its 
genuine static level (corresponding to $t\to\infty$, cf eqn~\ref{GLM}).  
Likewise, we only need the component with $\alpha'=\gamma'=1$ of the $C^{vv}_{\alpha'
\gamma'}$ 
tensor.  Thus we get using eqn~\ref{Ie18}:  
\[C^{\Delta\Delta}_{1111}=\frac {q^2}{\rho_0^2}\int_{-\infty}^{\infty}\frac {{\rm d}
\omega}{2\pi}\Delta\tilde {L}(q,s={\rm i}\omega +\epsilon )\Delta\tilde {L}(q,s=
-{\rm i}\omega +\epsilon )\]
\begin{equation}\cdot\left[2\pi\rho_0\delta (\omega )-q^2\left(\left(\frac {\tilde {
C}_L(q,s)}{s^2}\right)_{s={\rm i}\omega +\epsilon}+c.c.\right)\right]\label{Ie26}\end{equation}
where c.c. means complex conjugate and 
\begin{equation}\tilde {C}_L(q,s)\equiv\tilde {C}_{1111}(\underline q,s)=\frac {
\tilde {L}(q,s)}{1+\tilde {L}(q,s)q^2/\left(\rho_0s\right)}\label{Ie30}\end{equation}
is the Laplace transform of the longitudinal component of the stress 
correlation function (cf ref.~\cite{Polymers2024}).  

 Eqn~\ref{Ie26} can be simplified as (on using eqn~\ref{Ie30}): 
\begin{equation}C^{\Delta\Delta}_{1111}=\int_{-\infty}^{\infty}\frac {{\rm d}\omega}{
2\pi}\left|\Delta\tilde L(q,s={\rm i}\omega +\epsilon )\right|^{\,2}\left[\left(\frac 
1{\tilde {L}(q,s)+\rho_0s/q^2}\right)_{s={\rm i}\omega +\epsilon}+c.c.\right]\label{Ie32}\end{equation}
The latter expression can be further simplified by noting that 
both factors in the integrand are even functions of $\omega$ since $L(q,t)$ is 
real (at least for all isotropic achiral 
systems~\cite{SM2018,Polymers2024}), hence
\begin{equation}C^{\Delta\Delta}_{1111}=\int_{-\infty}^{\infty}\frac {{\rm d}\omega}{
\pi}\left|\Delta\tilde L(q,s={\rm i}\omega +\epsilon )\right|^{\,2}\left(\frac 1{
\tilde {L}(q,s)+\rho_0s/q^2}\right)_{s={\rm i}\omega +\epsilon}\label{Ie33}\end{equation}
The above equation is general. 
In order to get a more explicit result we need to specify the 
relaxation function $L(q,t)$. As a simple model we neglect the fast 
relaxation modes (due to particle vibrations and collisions) and assume  
\begin{equation}L(q,t)=L_e+A\exp\left(-t/\tau_{\alpha}\right),~L_0\equiv L(q,t=0
)=L_e+A\label{Ie34}\end{equation}
where all the three parameters $(L_e$, $A$ and $\tau_{\alpha}$) can depend on $q$.  Here 
$\tau_{\alpha}$ is considered as a terminal $(\alpha$-) relaxation time of the system, 
and we assume that this time is long enough (as compared to $1/\omega_s$, 
where $\omega_s=q\sqrt {L_0/\rho_0}$ is the sound frequency for the given $q$).   On 
using eqn~\ref{Ie34} we obtain 
\begin{equation}\Delta\tilde {L}(q,s)=\frac A{s+\delta},~\tilde {L}(q,s)=\frac {
L_e}s+\frac A{s+\delta}~\label{Ie35}\end{equation}
where $\delta =1/\tau_{\alpha}$. 
Eqn~\ref{Ie33},~\ref{Ie35} lead to
\begin{equation}C^{\Delta\Delta}_{1111}=\int_{-\infty}^{\infty}\frac {{\rm d}\omega}{
\pi}\frac {A^2}{\omega^2+\delta^2}\left(\frac s{L_e+As/(s+\delta )+\rho_0s^2/q^2}\right
)_{s={\rm i}\omega +\epsilon}\label{Ie36}\end{equation}
Recalling that $\delta$ is small,  
\begin{equation}\delta =1/\tau_{\alpha}\ll q\sqrt {L_0/\rho_0}\label{Ie38}\end{equation}
(which means that the relaxation time $\tau_{\alpha}$ is much longer than the 
time-period of acoustic waves with the given wave-number $q$) we can 
split the integral in eqn~\ref{Ie36} into two parts corresponding to 
$\left|s\right|\sim\delta$ (`singular' part $C_{sing}$) and the `regular' part $
C_{L0}$ accounting 
for $\left|s\right|\gg\delta$:  
\begin{equation}C_{L0}=\int_{-\infty}^{\infty}\frac {{\rm d}\omega}{\pi}\frac {A^
2}{\omega^2+\delta^2}\left(\frac s{L_0+\rho_0s^2/q^2}\right)_{s={\rm i}\omega +\epsilon}\label{Ie40}\end{equation}
and  
\begin{equation}C_{sing}\simeq\int_{-\infty}^{\infty}\frac {{\rm d}\omega}{\pi}\frac {
A^2}{\omega^2+\delta^2}\frac {A\delta}{L_0^2}\left(\frac s{s+\kappa\delta}\right
)_{s={\rm i}\omega +\epsilon}\label{Ie42}\end{equation}
where
\begin{equation}\kappa\equiv L_e/L_0\label{Ie43}\end{equation}
Evaluating the two above integrals we get
\begin{equation}C_{sing}\simeq\frac {A^3}{L_0^2}\frac 1{1+\kappa}~,~~~C_{L0}=\frac {
A^2}{L_0+\rho_0\delta^2/q^2}\label{Ie44}\end{equation}
Hence finally we obtain (taking also into account the condition of 
eqn~\ref{Ie38}): 
\begin{equation}C^{\Delta\Delta}_{1111}=C_{L0}+C_{sing}\simeq 2L_0\frac {(1-\kappa 
)^2}{1+\kappa}\label{Ie46}\end{equation}
As typically $\kappa <0.5$~(cf refs.~\cite{JCP164505,SoftM7867}) the 
above equation implies that $C^{\Delta\Delta}_{1111}$ is positive and not small, hence the 
stress noise and the reduced deviatoric stress must be really 
different, $\sigma^n_{11}\neq\sigma^{rd}_{11}$.  Indeed, the variance of $\sigma^{
\Delta}_{11}=\sigma_{11}^{rd}-\sigma_{11}^n$ at $t=0$ 
can be compared with the variances of $\sigma_{11}^{rd}$ or $\sigma_{11}^n$ which are equal to 
(see eqn~\ref{Ie17}) 
\begin{equation}M_{1111}(q,t=0)=L_0-L_e=L_0\left(1-\kappa\right)\label{Ie47}\end{equation}
Based on eqn~\ref{Ie46},~\ref{Ie47} we can conclude that both 
variances are comparable (for $\kappa\lesssim 0.5$), so that the difference $\sigma^{
\Delta}$ 
between $\sigma^{rd}$ and $\sigma^n$ at $t=0$ is significant and cannot be neglected.  
Indeed, the Pearson's correlation coefficient ($r$$^P$) for $\sigma_{11}^n(\underline 
q,t)$ and 
$\sigma_{11}^{rd}(\underline q,t)$ obtained based on eqn~\ref{Ie46},~\ref{Ie47} at $
t=0$ is 
\[r_{rd,n}^P=\frac {2\kappa}{1+\kappa}\]
For $\kappa\lesssim 0.5$ the correlation factor $r_{rd,n}^P$ is well below 1
meaning that the stress noise and the reduced deviatoric stress are 
far from being perfectly correlated at $t=0$.  

On the other hand, as follows from eqn~\ref{Ie14}, this difference 
decreases significantly at $t\gg\tau_{\alpha}$:  in this long-time regime the time 
argument in the kernel $M$ involved in the rhs of eqn~\ref{Ie14} is 
$t-t'>t\gg\tau_{\alpha}$, so that $M$ gets exponentially small, and the same must 
be true for the resultant $\sigma^{\Delta}$.  Therefore the two processes, $\sigma^{
rd}_{\alpha\beta}(\underline q,t)$ 
and $\sigma^n_{\alpha\beta}(\underline q,t)$, become asymptotically equal at $t\gg
\tau_{\alpha}$.

\subsection{\label{sI3} Discussion on stationarity of $\sigma^{rd}_{\alpha\beta}(\underline q,t)$ and $\sigma^n_{\alpha\beta}(\underline q,t)$} 

For liquids at equilibrium considered in the present paper the 
processes like $\sigma_{\alpha\beta}(\underline q,t),~v_{\gamma}(\underline q,t)
,~\rho (\underline q,t)$ are jointly stationary.  It 
means, in particular, that (auto- or cross-) correlation functions 
of these variables (like $\left<x(t_1)y(t_2)\right>$) depend only on the time-shift, 
$t_2-t_1$.  As $\sigma^d(\underline q,t)$ is related to the current $\rho_0v_{\gamma}
(\underline q,t)$, cf eqn~\ref{e7}, 
it is stationary as well, and the same is true for the stress noise 
$\sigma^n=\sigma -\sigma^d$ since the difference of two jointly stationary processes 
is also stationary.  By the same token $\sigma^{dev}=\sigma -\sigma^p$ is stationary 
because $\sigma^p(\underline q,t)$ is directly related to the mass density fluctuations, 
cf eqn~\ref{Ie6}.  Turning to the reduced deviatoric stress, $\sigma^{rd}_{\alpha
\beta}(\underline q,t)$, 
it can be shown rather straightforwardly based on its definition, 
eqn~\ref{Ie4a}, that its auto-correlation function 
\begin{equation}\frac VT\left<\sigma^{rd}_{\alpha\beta}(\underline q,t+t')\sigma^{
rd}_{\gamma\delta}(\underline q,t')^{*}\right>=M_{\alpha\beta\gamma\delta}(\underline 
q,t)\label{Mgen}\end{equation}
is independent of $t'$, ie it depends on the time shift $t$ only, cf 
eqn~\ref{Mdef}.  However, this feature does not mean that $\sigma^{rd}_{\alpha\beta}
(\underline q,t)$ 
is necessarily jointly stationary with other processes.  On the 
contrary, below we argue that the processes $\sigma^{rd}(\underline q,t)$ and $\sigma^
n(\underline q,t)$ are 
{\em not\/} jointly stationary.  Indeed, let's assume that $\sigma^{rd}(\underline 
q,t)$ and $\sigma^n(\underline q,t)$ 
are jointly stationary.  Then the difference $\sigma^{\Delta}(\underline q,t)=\sigma^{
rd}(\underline q,t)-\sigma^n(\underline q,t)$ 
must be stationary as well, so that its autocorrelation function 
$\left<\sigma^{\Delta}(\underline q,t)\sigma^{\Delta}(\underline q,t)^{*}\right>$ must be independent of time $
(t$).  However we know 
that at $t=0$ this tensorial function (more precisely, at least its 
longitudinal component) is positive, while at $t\gg\tau_{\alpha}$ this function 
must tend to zero.  We therefore arrive at a contradiction proving 
that $\sigma^{rd}(\underline q,t)$ and $\sigma^n(\underline q,t)$ are not jointly stationary.  

One may wonder how comes that $\sigma (\underline q,t)$, $\sigma^d(\underline q,
t)$, $\sigma^{dev}(\underline q,t)$, $\sigma^n(\underline q,t)$, 
etc are jointly stationary while adding $\sigma^{rd}(\underline q,t)$ to the list ruins this 
property.  This issue may be clarified as follows:  The point is that 
$\sigma (\underline q,t)$ comes from the classical Hamiltonian dynamics defined by the 
Liouvillean ${\cal L}$ involved in the time-evolution operator ${\cal R}(t)=e^{{\rm i}
{\cal L}t}$ (cf.  
refs.~\cite{FuchsPRL2017,FuchsJCP2018,SM2025}).  In this case the 
time-dependence of any variable $A$ is defined by the evolution of the 
system microstate $\Gamma =\Gamma (t)$:  $A(t)=A(\Gamma (t))$, where $\Gamma (t)
\equiv\Gamma (t,\Gamma_0)$ is the 
microstate occupied by the system at time $t$ provided that its initial 
microstate at $t=0$ was $\Gamma_0\equiv\Gamma (0)$.  As the ensemble of microstate 
trajectories, $\Gamma (t)$, is stationary in equilibrated Hamiltonian systems 
due to the uniformity of time, the same is true for any variable $A$ 
that is uniquely defined by the current microstate $\Gamma (t)=\Gamma (t,\Gamma_
0)$ 
(including $A=\sigma (\underline q$)).  That is why all variables mentioned above 
(except $\sigma^{rd}$) are jointly stationary since their sets at any time $t$ are 
unique functions of the current system microstate.  For example, 
recall that $\sigma_{\alpha\beta}^d(\underline q,t)$ is defined in eqn~\ref{e7} which can be written 
as 
\[\sigma^d_{\alpha\beta}(\underline q,t)=\int_0^{\infty}E_{\alpha\beta\gamma\delta}\left
(\underline q,\tau\right){\rm i}v_{\gamma}\left(\underline q,t-\tau\right)q_{\delta}
{\rm d}\tau\]
and therefore $\sigma_{\alpha\beta}^d(\underline q,t)$ can be considered as a unique function of $
\Gamma (t)$ 
since the velocity $v_{\gamma}\left(\underline q,t-\tau\right)$ is defined by the microstate $
\Gamma (t-\tau )$ 
which (for a given $\tau$) is in turn uniquely defined by $\Gamma (t)$ due to 
time-uniformity.   The same statement is also valid for $\sigma_{\alpha\beta}^n(\underline 
q,t)$ 
simply because $\sigma_{\alpha\beta}^n(\underline q,t)=\sigma_{\alpha\beta}(\underline 
q,t)-\sigma_{\alpha\beta}^d(\underline q,t)$ (see eqn~\ref{e6}).  

However the reduced deviatoric stress $\sigma^{rd}(\underline q,t)$ shows an exception 
from this rule since its time-dependence is defined by the artificial 
reduced evolution operator ${\cal R}'(t)$ (cf eqn~\ref{Rprime},~\ref{Ie4a}), so 
that $\sigma^{rd}(\underline q,t)$ becomes generally a time-dependent mixture of stress 
and momentum density, and therefore cannot be uniquely defined 
solely by the current state $\Gamma (t)$.  Indeed, on using 
eqn~\ref{rdn},~\ref{Ie14} we get 
\[\sigma^{rd}_{\alpha\beta}(\underline q,t)=\sigma_{\alpha\beta}^n(\underline q,
t)+\int_t^{\infty}M_{\alpha\beta\gamma\delta}(\underline q,\tau ){\rm i}v_{\gamma}
(\underline q,t-\tau )q_{\delta}{\rm d}\tau\]
where the integral in the rhs depends not only on $\Gamma (t)$ but also shows 
an additional $t$-dependence coming from the lower limit $\tau =t$.  
Therefore $\sigma^{rd}_{\alpha\beta}(\underline q,t)-\sigma_{\alpha\beta}^n(\underline 
q,t)$ is non-zero at $t=0$, but always vanishes 
at $t\to\infty$.  This behavior shows that $\sigma^{rd}_{\alpha\beta}(\underline 
q,t)-\sigma_{\alpha\beta}^n(\underline q,t)$ is not a 
stationary process, and hence the two processes, $\sigma^{rd}_{\alpha\beta}(\underline 
q,t)$ and 
$\sigma_{\alpha\beta}^n(\underline q,t)$, cannot be {\em jointly\/} stationary.  This feature clearly 
emphasizes the non-trivial nature of the exact general relations 
between the $M$-, $E$- and $C^n$-tensors (cf eqn~\ref{MCn},~\ref{f5}) valid 
for arbitrary time-shifts.  

\section{\label{DS}Discussion and Summary}
~

{\bf 1. Why the velocity field should be defined with mass factors.  }

In the present paper we used the mass factors to define the 
velocity, strain and concentration fields (cf eqs.~\ref{crt},~\ref{vrt}), 
which therefore generally differ from the so-to-say `conventional' (or 
`standard') definitions of these fields (not involving the mass 
factors).~\cite{HansDon,EvansMor} We mentioned this important 
definition problem at the end of section~\ref{Doem}.  Its more 
detailed discussion is presented below.  

There are several important points to emphasize concerning the 
relevance of particle masses to the strain and concentration fields, 
and eventually for the $q$-dependent relaxation moduli $E_{\alpha\beta\gamma\delta}
(\underline q,t)$:  
\vskip0.05in \noindent
{\bf (i)} According to the definition provided in eq.~\ref{e0} the moduli 
$E_{\alpha\beta\gamma\delta}(\underline q,t)$ do depend on how the velocity field, $
v_{\alpha}(\underline r,t)$, and the 
related strain rate, $\dot{\gamma}_{\alpha\beta}(\underline r,t)=\partial v_{\alpha}
(\underline r,t)/\partial r_{\beta}$, are defined.  Our point is 
that in the general case the velocity field must be defined in 
eq.~\ref{vrt} involving mass factors, $m_i/\bar {m}$, which are not involved 
in the `standard' definition of the velocity field, 
\begin{equation}\underline v^{st}(\underline r,t)=\frac 1{c_0}\sum_i\underline v_
i(t)\delta (\underline r-\underline r_i(t))\label{vrtst}\end{equation}
and the microscopic concentration,
\begin{equation}c^{st}(\underline r,t)=\sum_i\delta (\underline r-\underline r_i
(t))\label{crtst}\end{equation}
The above equations obviously become equivalent to 
eqn~\ref{vrt},~\ref{crt} in the case of equal mass of all particles.  
The most important arguments supporting our view that in the 
general case of different particle masses, $m_i$, the definitions, 
eqn~\ref{crt},~\ref{vrt}, must be favored are presented below:  

$\bullet$ Let us turn to a polydisperse oligomer melt~\cite{RubColby} 
considering each chain as a `particle'.  In this case eq.~\ref{crtst} 
would provide the chain-number concentration field.  Is it appropriate 
to define a generalized $q$-dependent longitudinal modulus $L(q,t)$ in 
terms of such a concentration field?  The answer is probably `no', 
given that $c^{st}(\underline r,t)$ is expected to strongly fluctuate as one chain of, 
say, 10 units may be replaced by 2 chains of 5 units thus increasing 
the local chain number concentration $c^{st}$ by a factor of $\sim 2$.  It 
seems clear that such a strong fluctuation should not be related to 
the compressibility of the system.  Such a problem is avoided if 
the contribution of each chain is scaled with its mass, thus leading 
to the definitions of $c(\underline r,t)$ and $\underline v(\underline r,t)$ adopted in the present paper 
(see eqs.~\ref{crt},~\ref{vrt}).  

$\bullet$ Another way to come to the same conclusion is outlined below:  
Consider a mixture of small spherical beads (unimers of mass $m$) and 
dimers (two beads of total mass $2m$ connected by an ideal spring with 
elastic constant $k$).  At low $k$ the beads in a dimer are far apart, so 
that it is natural to assume that both velocities of the dimer beads 
must be counted on the equal basis with a unimer bead in the 
collective velocity $\underline v(\underline r,t)$, which is in accordance with eq.~\ref{vrt} 
we adopted.  As the constant $k$ increases, the beads of a dimer 
become closer to each other, but there is no reason for the dimer 
velocity to suddenly become weighted twice lighter at some value of 
$k$, hence our definition, eq.~\ref{vrt}, is likely to be always 
appropriate.  

\vskip0.05in \noindent\  
{\bf (ii)} The dynamics of supercooled glass-forming liquids and other 
complex fluids (including flow, deformation and sound waves) are 
largely defined by the microscopic internal stress field, $\sigma_{\alpha\beta}$, as well 
as the mass density, $\rho$, and mass current density, $\underline J=\rho_0\underline 
v$.  
Importantly, all these fields are related by the general conservation 
laws, cf eqs.~\ref{mbal},~\ref{momeq}.  It is therefore natural to 
define the elastic moduli in terms of these fields:  in this case the 
important relations between the moduli and correlation functions of 
microscopic fields (like eqn~\ref{e18} known from the classical theory 
developed for monodisperse systems~\cite{EvansMor}) would be valid 
also in the more general case of mass polydispersity.  Hence, for 
example, the strain rate field must be defined in terms of the 
generalized collective velocity proportional to the momentum density, 
in agreement with eqn~\ref{vrt} adopted in the present paper (cf.  
eqs.~\ref{vrt},~\ref{crt} and 
refs.~\cite{JCP164505,Polymers2024,SM2025}).  To reiterate this point, 
the classical theories of fluid dynamics actually follow the ideas 
mentioned above~\cite{BalZop,EvansMor,Forster,HansDon}, being, 
however, mostly focussed on the case of particles with equal mass, 
and, perhaps, sometimes making a wrong impression that the 
definitions like eq.~\ref{vrtst} are more general.  

It is also important to stress that in order to impose a generalized 
(non-uniform) deformation on an amorphous system it is necessary to 
apply an appropriate external force field acting on all 
particles.~\cite{JCP164505,Polymers2024} In the case of a deformation 
wave with wavevector $\underline q$ the force on a particle $`i$' is:  
\begin{equation}\underline F_i(t)=m_i\underline A(t)e^{{\rm i}\underline q\cdot\underline 
r_i(t)}+c.c.\label{eqF}\end{equation}
where c.c.  means complex conjugate.  The factors $m_i$ in the above 
equation are providing the coherent acceleration of all particles.  
Noteworthily, the standard definitions of the relaxation moduli, 
$E_{\alpha\beta\gamma\delta}(\underline q,t)$, are hinged on the instantaneous canonical transformations 
in the Hamiltonian phase space~\cite{Lutsko,MolPhys2881,SM2018} 
whose effect can be exactly reproduced by application of appropriate 
perturbative external forces involving the very mass factors that are 
present in eq.~\ref{eqF}.~\cite{Polymers2024,SM2018} 

\vskip0.05in \noindent
{\bf (iii) Velocity field and energy.}  As argued above a natural way to 
create a deformation is to apply an external force $\underline F_i(t)$ defined in 
eq.~\ref{eqF}, which leads to the following rate-of-change of the total 
energy $H$:  
\begin{equation}\frac {{\rm d}H}{{\rm d}t}=\sum_i\underline F_i(t)\cdot\underline 
v_i(t)=V\underline A(t)\cdot\underline J(-\underline q,t)+c.c.\label{eqH}\end{equation}
(cf eq.~\ref{Jqt}).  The above equation shows that if the strain rate 
$\dot{\gamma}_{\alpha\beta}(\underline q,t)$ is defined in eq.~\ref{gammadot} using the microscopic 
velocity field of eqs.~\ref{vrt},~\ref{vqt}, the no-strain condition, 
$\dot{\gamma}_{\alpha\beta}(\underline q,t)=0$, would automatically imply that $
H={\rm c}{\rm o}{\rm n}{\rm s}{\rm t}\,$ as it should 
be in the case of no strain.  By contrast, if the `conventional' 
definition of the velocity field, eq.~\ref{vrtst}, is used instead (ie the 
strain rates are defined as $\dot{\gamma}_{\alpha\beta}^{st}(\underline q,t)={\rm i}
q_{\beta}v_{\alpha}^{st}(\underline q,t)$), the energy change 
rate, ${\rm d}H/{\rm d}t$, would not necessarily vanish for systems with mass 
polydispersity, which would mean that the applied force field can 
produce a non-zero work at zero strain (ie at $\dot{\gamma}_{\alpha\beta}^{st}(\underline 
q,t)=0$), which 
cannot be the case (as otherwise it would allow for a force applied 
to deform a solid system to produce a non-zero work at no 
deformation).  

\vskip0.1in \noindent

{\bf 2. Why the mass polydispersity is important for the static elasticity }
{\bf at} $q\neq 0$

A referee of this paper expressed an opinion that particle masses are 
irrelevant for the static properties of a liquid.  This is true, for 
example, for the bulk isothermal compressibility (associated with the 
pressure response to a uniform affine compression of the system, cf 
eq.~\ref{IIe5}).  However, such a statement is not applicable to 
$q$-dependent properties of {\em polydisperse\/} systems:  a mass 
polydispersity (as well as other kinds of particle 
dispersity~\cite{JCP164505}) can lead to a strong impact also on the 
static $q$-dependent moduli even if the polydispersity index (PDI) is 
low (like $\mbox{\rm PDI}\sim 1\%$)~\cite{PRE042611} (in contrast to the bulk static 
moduli).  

The important point here is that in order to obtain a generalized 
$q$-dependent modulus (eg $L(q,t)$ or $M(q,t)$) it is necessary to create a 
deformation {\em wave}, and this cannot be done by applying pressure only 
at the surface of the system (like in the case of uniform 
compression).  Instead, it is necessary to apply an appropriate 
external force $\underline F_i(t)$ to each particle $`i$' (see eq.~\ref{eqF}) which is 
proportional to its mass  (see point 1(ii) above).  

It is also very important that the general case of the particle mass 
and/or size polydispersity, the multi-component nature of the system 
leads to a slow inter-diffusion of particles according to their 
mass/size:  in spite of the strain wave being kept constant, the 
small and large (or light and heavy) particles tend to partially 
separate and redistribute~\cite{PRE042611} This process in the case of 
size polydispersity was elucidated in detail in ref.~\cite{JCP164505} 
(see section II.F there).  In this case, the initially homogeneous local 
{\em composition\/} of the system slowly renders harmonically modulated 
since large particles tend to escape from denser regions, being 
replaced there by smaller (and/or heavier) particles which penetrate 
easier into regions of higher pressure~\cite{JCP164505}.  As a result 
the {\em equilibrium\/} (ie genuinely static) elastic moduli, $L_e(q)$ and $M_e(
q)$, 
become dependent on the particle mass/size distribution.  

To reiterate the point on the {\em bulk\/} moduli which are hinged on {\em affine }
deformations:  in this case no external force is needed to maintain a 
constant affine strain, and that is why, say, the modulus $L_e^{bulk}$ (see 
eq.~\ref{IIe5}) does not depend on particle masses ($m_i$) in contrast to 
$L_e(q)$ which is generally different from $L_e^{bulk}$ in polydisperse systems 
% (including the hydrodynamic limit: $\lim_{q\to 0}L_e(q)<L_e^{bulk}$)
as explained in ref.~\cite{JCP164505} (and as mentioned below 
eq.~\ref{IIe5}).  

\vskip0.1in \noindent

{\bf 3.}  As was established in ref.~\cite{Polymers2024}, the equilibrium 
elasticity tensor $E^e_{\alpha\beta\alpha'\beta'}(\underline q)$ (which represents the genuinely static 
limit of the tensor $E_{\alpha\beta\alpha'\beta'}(\underline q,t)$ of generalized relaxation moduli) can 
be expressed in terms of two elastic moduli, longitudinal $L_e(q)$ 
and transverse $M_e(q)$.  In section~\ref{sII} we formulated a new 
approach allowing to obtain these equilibrium moduli 
based on structural correlation properties of the system (ie on 
correlations taken at the same time).   The main results are 
eqn~\ref{Legen} and~\ref{eMe} defining the moduli in terms of several 
structure factors.  This approach opens up a way for a faster and 
more precise calculation of the equilibrium moduli using MD 
simulations (cf section~\ref{Doem}).  It is worth emphasizing that 
eqn~\ref{Legen} and~\ref{eMe} are generally valid also for 
polydisperse systems (as discussed in section~\ref{Doem}).  

\vskip0.1in \noindent

{\bf 4.}  In the present paper we also derived and discussed the basic 
general relations between the elasticity tensor ${\bf E}(\underline q,t)$ resolved in 
space-time, and the stress correlation function ${\bf C}(\underline q,t)$ for equilibrium 
amorphous systems.  We employed the FDT result, eq.~\ref{f5}, to 
elaborate a conceptually new derivation of the basic relation, 
eqn~\ref{e2}, which does not involve consideration of either 
constrained dynamics or non-steady conditions (originally employed in 
ref.~\cite{Polymers2024} to obtain this relation).  Noteworthily, the 
general derivation presented in section~\ref{sIII} implies that the 
resultant eqn~\ref{e2} is applicable also to polydisperse systems with 
any kind of polydispersity and with a finite thermal conductivity 
(including extreme cases of isothermic and adiabatic conditions).  Such 
generalizations have not been developed as yet in the framework of 
the Zwanzig--Mori (ZM) projection operator formalism.  

\vskip0.1in \noindent

{\bf 5.}  It is worth mentioning that eqn~\ref{e2} was recently derived by 
an independent method using a rigorous decomposition of the classical 
dynamics via the ZM projection operator formalism~\cite{SM2025}.  
The ZM approach expresses $C_{\alpha\beta\alpha'\beta'}(\underline q,t)$ in terms of the projected 
stress-memory kernel~\cite{FuchsJCP2018,FuchsPRL2017}, $M_{\alpha\beta\alpha'\beta'}
(\underline q,t)$, 
which can be also considered as the generalized viscosity 
tensor~\cite{SM2025}.  Its time dependence is not generated by the 
classical Newtonian dynamics, but rather by an abstract `projected' 
dynamics which is not directly accessible in molecular dynamics 
simulations.  This obstacle had greatly hampered the development of 
accurate approximations for $M_{\alpha\beta\alpha'\beta'}(\underline q,t)$ in the past~\cite{Jan2}.  
Therefore, the intriguing findings of refs.~\cite{Polymers2024,SM2025} 
that the kernel ${\bf M}(\underline q,t)$ coincides with the stress noise correlation 
function ${\bf C}^n(\underline q,t)$ (cf.  eqn~\ref{MCn}), and that ${\bf C}^n(\underline 
q,t)$ is related to 
the elasticity tensor ${\bf E}(\underline q,t)$, which are reiterated in the present paper 
(cf.  eqn~\ref{f5}), 
\begin{equation}M_{\alpha\beta\alpha'\beta'}(\underline q,t)=C_{\alpha\beta\alpha'
\beta'}^n(\underline q,t)=E_{\alpha\beta\alpha'\beta'}(\underline q,|t|)-E_{\alpha
\beta\alpha'\beta'}^e(\underline q)\label{M2Cn}\end{equation}
are of primary importance.  
   
\vskip0.1in \noindent

{\bf 6.}  To continue from the previous point, the memory kernel ${\bf M}(\underline 
q,t)$ 
involved in eqn~\ref{e2} plays a central role in the Zwanzig-Mori 
projection operator formalism.~\cite{HansDon,Goetze,BalZop,Zwan,Mori} 
An important aspect here is that ${\bf M}(\underline q,t)$ is equal to the correlation 
function of the reduced deviatoric stress $\sigma^{rd}$ (cf eqn~\ref{Mdef}) 
generated by the non-Newtonian projected 
dynamics.~\cite{Goetze,FuchsJCP2018,FuchsPRL2017,SM2025} The relation 
${\bf M}(\underline q,t)={\bf C}^n(\underline q,t)$ (eqn~\ref{MCn}) therefore hints at a close 
relationship between the stress noise $\sigma^n(\underline q,t)$ and the reduced 
deviatoric stress $\sigma^{rd}(\underline q,t)$.  It has been shown~\cite{SM2025} that 
$\sigma^n(\underline q,t)=\sigma^{rd}(\underline q,t)$ in a special case when the momentum density at 
wavevector $\underline q$ is suppressed at $t<0$ by an appropriate external 
force, while this no-flow constraint is released at $t>0$.  In the 
present paper we demonstrated however that in the general case 
$\sigma^n(\underline q,t)$ is different from $\sigma^{rd}(\underline q,t)$ (see section~\ref{sI}) in spite of the 
fact that their autocorrelation functions are equal to each other.  
However, the two processes become asymptotically equal at long 
times, $t\gg\tau_{\alpha}$ (cf section~\ref{sI}).  Moreover, we argue that while 
the process $\sigma^n(\underline q,t)$ is generally stationary (in equilibrated systems), 
the difference $\sigma^{rd}(\underline q,t)-\sigma^n(\underline q,t)$ is not, showing that $
\sigma^{rd}(\underline q,t)$ and 
$\sigma^n(\underline q,t)$ are not jointly stationary and pointing to a nontrivial 
relationship between the Newtonian and the ZM-projected dynamics.  

\vskip0.25in \noindent

{\bf Data availability}
\vskip0.05in \noindent
The data that support the findings of this study are available from 
the corresponding author upon reasonable request.  All the relevant 
data generated during this study have been included in this published 
article.  

\vskip0.25in \noindent

{\bf Conflicts of interest}
\vskip0.05in \noindent
There are no conflicts to declare.

\vskip0.25in \noindent

%\acknowledgments
{\bf Acknowledgments}

This paper is dedicated to Prof.  Michael Rubinstein on the occasion 
of his 70th birthday.  We acknowledge funding from the French 
National Agency for Research (ANR-24-CE30-7668 `VMFCMD').

\appendix

\section{On the dissipation rate and its relation to the
stress noise} 

Let us consider a liquid near the equilibrium state, which is 
perturbed due to a weak external force equal to $\underline f_{ext}=\rho_0\underline 
A(\underline r,t)+c.c$.  
per volume, where the acceleration $\underline A(\underline r,t)$ is harmonic in space and 
time 
\[\underline A(\underline r,t)=\underline {\tilde {A}}\exp({\rm i}\underline q\cdot\underline 
r+{\rm i}\omega t)\]
(cf. eqn~\ref{eqF}). Such an external force generates a weak flow with 
velocity
\[\underline v(\underline r,t)=\underline {\tilde {v}}\exp({\rm i}\underline q\cdot\underline 
r+{\rm i}\omega t)+c.c.\]
where $\underline {\tilde {A}}$ and $\underline {\tilde {v}}$ are complex vectors. 
The external force and flow velocity are related due to the 
momentum equation (cf eqn~\ref{e11})
\begin{equation}{\rm i}\omega\tilde {v}_{\alpha}=\tilde {A}_{\alpha}+\tilde{\sigma}_{
\alpha\beta}{\rm i}q_{\beta}/\rho_0\label{af3}\end{equation}
where $\sigma_{\alpha\beta}(\underline r,t)=\tilde{\sigma}_{\alpha\beta}\exp({\rm i}\underline 
q\cdot\underline r+{\rm i}\omega t)+c.c.$ is the mean (ensemble 
averaged) stress field, which is therefore deterministic, 
$\sigma_{\alpha\beta}(\underline r,t)=\sigma_{\alpha\beta}^d(\underline r,t)$.  Note that `tilde' in the above equations indicates 
complex amplitudes (corresponding to the selected $\underline q$ and $\omega$), and that `tilde' is 
omitted in what follows.  

The dissipation rate ${\cal D}$ is equal to the work per time (and per 
volume) performed by the external force:  
\begin{equation}{\cal D}=\rho_0\underline A\cdot\underline v^{*}+c.c.\label{af4}\end{equation}
Using eqn~\ref{af3} the acceleration $\underline A$ 
can be expressed in terms of $\underline v$ and $\sigma_{\alpha\beta}=\sigma^d_{
\alpha\beta}$:  
\begin{equation}A_{\alpha}={\rm i}\omega v_{\alpha}-\sigma_{\alpha\beta}{\rm i}q_{
\beta}/\rho_0\label{afn}\end{equation}
The $\underline v$ contribution to $\underline A$ in the above equation does not give rise to 
any dissipation (by virtue of eqn~\ref{af4}), while the second term in 
the rhs of eqn~\ref{afn} involving $\sigma_{\alpha\beta}$ can be again expressed in terms 
of the flow velocity field (cf.  eqn~\ref{e7}):  
\[\sigma_{\alpha\beta}=E_{\alpha\beta\gamma\delta}(\underline q,\omega ){\rm i}v_{
\gamma}q_{\delta}\]
where $E_{\alpha\beta\gamma\delta}(\underline q,\omega )$ is the Fourier transform (cf eqn~\ref{e10}) of the 
tensor function $E_{\alpha\beta\gamma\delta}(\underline q,t)$ of the generalized relaxation moduli.  Thus 
we get 
\begin{equation}{\cal D}=E_{\alpha\beta\gamma\delta}(\underline q,\omega )q_{\beta}
q_{\delta}v_{\alpha}^{*}v_{\gamma}+c.c.\label{af5}\end{equation}
According to eqn~\ref{f5} the Fourier transform (eqn~\ref{e10}) of the 
$E$-tensor can be represented as (for $\epsilon\to 0$):  
\begin{equation}E_{\alpha\beta\gamma\delta}(\underline q,\omega )=\frac 1{{\rm i}
\omega}E^e_{\alpha\beta\gamma\delta}(\underline q)+\int_0^{\infty}C^n_{\alpha\beta
\gamma\delta}(\underline q,t)e^{-{\rm i}\omega t}{\rm d}t\label{af6}\end{equation}
It is straightforward to show that only the real part of $E$ does 
contribute to ${\cal D}$ in eqn~\ref{af5}.  Therefore we get (taking into 
account that $C^n$ is even in time):  
\begin{equation}{\cal D}=C^n_{\alpha\beta\gamma\delta}(\underline q,\omega )q_{\beta}
q_{\delta}v_{\alpha}^{*}v_{\gamma}\label{af7}\end{equation}
Finally, on using eqn~\ref{e15} and defining 
\begin{equation}X(\underline q,\omega )\equiv\sigma^n_{\alpha\beta}(\underline q
,\omega )v^{*}_{\alpha}q_{\beta}\label{af8}\end{equation}
we obtain the dissipation rate 
\begin{equation}{\cal D}=\frac {V\epsilon}T\left<X(\underline q,\omega )X^{*}(\underline 
q,\omega )\right>\label{af10}\end{equation}
which is obviously necessarily non-negative.  Note that ${\cal D}$ stays finite 
in the limit $\epsilon\to 0$ where $\epsilon =1/\Delta t$ (cf eqn~\ref{e15} and the sentence 
below it).

\section{Derivation of the generalized compressibility equation based 
on cross-correlations of stress and concentration} 

The longitudinal modulus $L_e(q)$ defines the static response of the 
longitudinal stress $\delta\sigma_{11}(\underline q)$ to a weak permanent (time independent) 
strain, $\delta\gamma (\underline q)$,~\cite{Polymers2024,SM2025} applied to the fully 
equilibrated system at $t=-\infty$,~\footnote{Note that here and below 
$q\neq 0$ and we use the naturally rotated coordinates (NRC) with axis $1$ 
parallel to the wavevector $\underline q$~\cite{SM2018,PRE2023,Polymers2024}.} 
\begin{equation}\delta\gamma (\underline q)\equiv\delta\gamma_{11}(\underline q)
={\rm i}q\delta u_1(\underline q)\label{gammad}\end{equation}
where {$\delta u_1(\underline q)$ is the amplitude of the harmonic field of particle
displacements parallel to $\underline q$, $\delta u_1(\underline r)=\delta u_1(\underline 
q)e^{{\rm i}\underline q\cdot\underline r}+c.c.$, applied to each 
system of a well-equilibrated canonical 
ensemble~\cite{Polymers2024}.} (Note that the strain tensor in the 
real space is defined as $\gamma_{\alpha\beta}=\frac {\partial u_{\alpha}}{\partial 
r_{\beta}}$.)  Thus, the {ensemble-averaged }
stress response is (see eqn~\ref{e0}, also cf ref.~\cite{Polymers2024} 
and eqn (31), (32) in ref.~\cite{SM2025}) 
\begin{equation}\delta\sigma_{11}(\underline q)=L_e(q)\delta\gamma (\underline q
)\label{Ledef}\end{equation}
  
Recalling that for weak deformations the rate-of-change of the 
collective displacement is equal to the collective velocity $\underline v(\underline 
q,t)$, 
\[\frac {\partial\underline u(\underline q,t)}{\partial t}=\underline v(\underline 
q,t)\]
and using eqn~\ref{mqbal} we get 
\begin{equation}\delta c(\underline q)/c_0=-{\rm i}q\delta u_1(\underline q)=-\delta
\gamma (\underline q)\label{cdef}\end{equation}
where $\delta c(\underline q)$ is the concentration change generated by a weak 
longitudinal collective displacement $u_1(\underline q)$. Using the above equation we 
can rewrite eqn~\ref{Ledef} as 
\begin{equation}\delta\sigma_{11}(\underline q)=-L_e(q)\delta c(\underline q)/c_
0\label{Lec}\end{equation}
which means that $L_e(q)$ also defines the static longitudinal stress 
response to a concentration perturbation.  

Noteworthily, a permanent concentration perturbation can be created 
by a weak static external field $U_{ext}(\underline r)$ applied to each particle and 
associated with a perturbative term in the total Hamiltonian 
$({\cal H}={\cal H}_0+\Delta {\cal H}$) of the system:  
\begin{equation}\Delta {\cal H}=-U_0c^{*}(\underline q)\label{dHc}\end{equation}
(Note that here and below we consider the uniform equilibrium state 
where the mean $c(\underline q)$ vanishes as the reference undeformed ensemble.  
Hence, $c(\underline q)$ and $\delta c(\underline q)$ are always the same upon application of a 
strain.)  The above equation can be written as (cf eqn~\ref{cqt}) 
\[\Delta {\cal H}=\sum_iU_{ext}^{(i)}(\underline r_i)\]
where
\begin{equation}U_{ext}^{(i)}(\underline r)=-\frac {m_i}{\bar {m}V}U_0e^{{\rm i}\underline 
q\cdot\underline r}\label{Uext}\end{equation}
is the external field applied to particle $i$.  As we consider the static 
case, the mean (time-averaged) force on each volume element must be 
zero, hence by virtue of the momentum balance equation 
\[\frac {\partial\sigma_{\alpha\beta}}{\partial r_{\beta}}+f_{\alpha}^{ext}=0\]
where 
\[f_{\alpha}^{ext}(\underline r)=-\frac 1V\frac {\partial}{\partial r_{\alpha}}\sum_
iU_{ext}^{(i)}(\underline r)=\frac 1Vc_0U_0{\rm i}q_{\alpha}e^{{\rm i}\underline 
q\cdot\underline r}\]
is the ensemble-averaged external force per volume (note that the 
prefactor $1/V$ in the above equation is the probability density that a 
particle $i$ is present near the position $\underline r$).  Based on the above two 
equations for $\alpha =1$ (longitudinal component) we get the time-averaged 
perturbation of the stress $\sigma_{11}$:  
\begin{equation}\delta\sigma_{11}(\underline r)=-c_0\frac {U_0}Ve^{{\rm i}\underline 
q\cdot\underline r},~\delta\sigma_{11}(\underline q)=-c_0\frac {U_0}V\label{ee2}\end{equation}
Furthermore, using the fluctuation-dissipation theorem (FDT) saying 
that the correlation function of two variables $A$ and $B$ (both with 
zero mean values) is proportional to the response of $A$ to the 
external field conjugate to $B^{*}$, we get (considering $A=\sigma_{11}(\underline 
q)$,  
$B=c(\underline q$) and recalling eqn~\ref{dHc}):  
\begin{equation}\delta\sigma_{11}(\underline q)=\frac {U_0}T\left<\sigma_{11}(\underline 
q)c^{*}(\underline q)\right>\label{ee4}\end{equation}
Then based on the above equation and eqn~\ref{ee2},~\ref{ee4} we 
finally obtain 
\begin{equation}\left<\sigma_{11}(\underline q)c^{*}(\underline q)\right>=-\frac {
c_0T}V\label{ee6}\end{equation}
It is worth noting that eqn~\ref{ee6} is valid also for polydisperse 
systems including those with mass polydispersity (provided that the 
local concentration is defined in eqn~\ref{crt} involving 
mass-dependent weight factors).  

Furthermore, noting that in the general case we can always write
\begin{equation}A=\lambda B+X,\label{eAB}\end{equation}
where $\lambda$ is an unknown number and $X$ is independent of $B$, $\left<XB^{*}\right
>=0$, 
$\left<XB\right>=0$ (note that $\left<XB\right>=0$, $\left<BB\right>=0$ are valid since both $
X$ and $B$ 
correspond to a non-zero $\underline q$ and the equilibrium system we consider is 
macroscopically uniform).  Taking into account that for a large 
system the variables $A$ and $B$ are small fluctuations which are 
jointly Gaussian (and therefore the same is true for $X$) we conclude 
(based on the above 
equations) that the conditional average of $X$ for any prescribed 
value of $B$, $\left<X\right>_B$, always vanishes.  Hence $\lambda$ is the ratio of two 
correlators, 
\begin{equation}\lambda =\frac {\left<AB^{*}\right>}{\left<BB^{*}\right>}\label{lam}\end{equation}
For $A=\sigma_{11}(\underline q)$ and $B=c(\underline q$), based on 
eqn~\ref{Lec},~\ref{eAB} we therefore identify $\lambda$ as 
\begin{equation}\lambda =-L_e(q)/c_0\label{lam2}\end{equation}
Furthermore, in this case
\begin{equation}\left<BB^{*}\right>=\left<c(\underline q)c^{*}(\underline q)\right
>=\frac N{V^2}S_2(q)\label{cqcq}\end{equation}
where $S_2(q)$ is the structure factor defined in the general case of 
particle mass polydispersity in eq.~\ref{Sqpoly} which is 
different from the standard structure factor $S(q)$ (cf eq.~\ref{eSq}). 

Using eqn~\ref{ee6},~\ref{lam},~\ref{lam2},~and \ref{cqcq} we finally 
get the generalized compressibility equation (cf eqn~\ref{Legen}):  
\[L_e(q)=c_0T/S_2(q)\]
This result agrees with the theory presented  in 
ref.~\cite{Polymers2024}.


\begin{thebibliography}{999}

\bibitem{RubColby} M.Rubinstein and R.H.Colby, {\em Polymer Physics}, 
Oxford U Press, NY, 2003.  

\bibitem{Bird} R.B. Bird, R.C.  Armstrong and O.  Hassager, {\em Dynamics of }
{\em Polymeric Liquids:  Fluid Mechanics\/}; Wiley, NY, 1987.  

\bibitem{Heuer} A.Heuer, J.Phys.: Cond.Matt., 2008, {\bf 20}, 373101.

\bibitem{Ferry} J.D.  Ferry, {\em Viscoelastic Properties of Polymers}, 3rd 
ed.,Wiley:  New York, NY, 1980.  

\bibitem{Nicolas} A. Nicolas,  E.E. Ferrero, K. Martens and  J.-L. Barrat, 
Rev.  Mod.  Phys.,  2018, {\bf 90}, 045006.  

\bibitem{Fuchs} M. Fuchs, Adv. Polym. Sci., 2010, {\bf 236}, 55.

\bibitem{Berthier} L.  Berthier and G.  Biroli, Rev.  Mod.  Phys.,  2011, 
{\bf 83}, 587.  

\bibitem{JCP164505} L.  Klochko, J.  Baschnagel, J.P.  Wittmer, H.  
Meyer, O.  Benzerara and A.N.  Semenov, J.  Chem.  Phys.,  2022, {\bf 156}, 
164505.  

\bibitem{microR} Q.Xia, H.Xiao, Y.Pan and L.Wang, Adv.Coll.Int.Sci., 
2018, {\bf 257}, 71.  

\bibitem{Polymers2024} A. N. Semenov and J. Baschnagel, Polymers, 2024, {\bf 16},
2336.

\bibitem{SM2025} N.Grimm et al., Soft Matter, 2025, {\bf 21}, 4256.  

\bibitem{SoftM7867} L.Klochko, J.Baschnagel, J.P.Wittmer and 
A.N.Semenov, Soft Matter, 2021, {\bf 17}, 7867.  

\bibitem{SM2018} L.Klochko, J.Baschnagel, J.P.Wittmer and 
A.N.Semenov, Soft Matter, 2018, {\bf 14}, 6835.  

\bibitem{EvansMor} D.J.Evans and G.Morriss, {\em Statistical Mechanics }
{\em of Nonequilibrium Liquids}, Cambridge University Press, London, 2008.  

\bibitem{FuchsPRL2017} M.Maier, A.Zippelius and M.Fuchs, Phys.  Rev.  
Lett. 2017, {\bf119}, 265701.  

\bibitem{FuchsJCP2018} M.  Maier, A.  Zippelius and M.  Fuchs, J.  
Chem.  Phys.,  2018, {\bf 149}, 084502.  
 
\bibitem{HansDon}J.  P.  Hansen and I.  R.  McDonald, {\em Theory of }
{\em Simple Liquids}, Academic Press, London, 1986.  

\bibitem{BalZop}U.  Balucani and M.  Zoppi, {\em Dynamics of the Liquid }
{\em State}, Oxford University Press, Oxford, 2003.  

\bibitem{Gold2002} I.  Goldhirsch and C.  Goldenberg, Eur.  Phys.  J.  
E, 2002, {\bf 9}, 245.  

\bibitem{Shi2023} K.Shi, E.R.Smith, E.E.Santiso, K.E.Gubbins, 
J.Chem.Phys., 2023, {\bf 158}, 040901.

\bibitem{LL5} L.D.Landau and E.M.Lifshitz, {\em Statistical physics}, v.5, 
Elsevier, Amsterdam, 2005.

\bibitem{LL6} L.D.Landau and E.M.Lifshitz, {\em Fluid Mechanics}, v.6, 
Pergamon Press, Oxford, 1986.  

\bibitem{Leq} Y.Pomeau and J.Piasecki, C.R.Physique, 2017, {\bf 18}, 570.  

\bibitem{ad1} J. Farago, H. Meyer and A. N. Semenov,
Phys.Rev.Lett., 2011, {\bf 107}, 178301.

\bibitem{FaragoPRE051807} J. Farago, H. Meyer, J. Baschnagel and A.N. Semenov,
Phys. Rev. E, 2012, {\bf 85}, 051807.

\bibitem{SemJCP244905} A.  N.  Semenov, J.  Farago and H.  Meyer, 
J.  Chem.  Phys., 2012, {\bf 136}, 244905.  

\bibitem{ad7} H. Meyer and A.N. Semenov, Phys. Rev. Lett, 2012, {\bf 109}, 248304.

\bibitem{ad8} A.  N.  Semenov and H.  Meyer, Soft Matter, 2013, {\bf 9}, 
4249.  

\bibitem{PRE2023} J.  P.  Wittmer, A.  N.  Semenov, and J.  
Baschnagel,   
Phys. Rev. E, 2023, 108, 015002.  

\bibitem{swaps} A.  Ninarello, L.  Berthier and D.  Coslovich, Phys.  
Rev.  X, 2017, 7, 021039.  

\bibitem{Goetze} W.  G\"otze, {\em Complex Dynamics of Glass-Forming }
{\em Liquids:  A Mode-Coupling Theory}, Oxford University Press, Oxford, 
2009.  
 
\bibitem{P4} L.M.C. Janssen, Front.  Phys.,  2018, {\bf 6}, 97.  

\bibitem{P5} L.  Berthier and G.  Biroli, Rev.  Mod.  Phys.,  2011, {\bf 83}, 
587.  

\bibitem{Jan8} C.  Luo and L.  M.  C.  Janssen, J.  Chem.  Phys.,  
2020, {\bf 153}, 214506.  

\bibitem{Jan10} F.  Weysser, A.  M.  Puertas, M.  Fuchs and Th.  
Voigtmann, Phys.  Rev.  E, 2010, {\bf 82}, 011504.  

\bibitem{Voronoi} C.  Ruscher, A.  N.  Semenov, J.  Baschnagel, and J.  
Farago, J. Chem. Phys. 2017, {\bf 146}, 144502. 

\bibitem{OZ}L.  S.  Ornstein and F.  Zernike, Z.  Physik, 1918, {\bf 19}, 134.  

\bibitem{PRE042611} L.  Klochko, J.  Baschnagel, J.  P.  Wittmer, O.  
Benzerara, C.  Ruscher and A.  N.  Semenov, Phys.Rev.E, 2020, {\bf 102}, 
042611.  

\bibitem{Jan2} I.  Pihlajamaa, V.  E.  Debets, C.  C.  L.  Laudicina and 
L.  M.  C.  Janssen, SciPost Phys.,  2023, {\bf 15}, 217.  

\bibitem{Jan7} M.  Kerr Winter, I.  Pihlajamaa, V.  E.  Debets, and L.  
M.  C.  Janssen,  J.Chem.Phys., 2023, {\bf 158}, 244115.  
 
\bibitem{MolPhys2881} J.P.Wittmer, H.Xu, O.Benzerara and J.Baschnagel, 
Mol.Phys.,  2015, {\bf 113}, 2881.  

\bibitem{RR} I.Rubinstein and L.Rubinstein, {\em Partial differential}
{\em equations in classical mathematical physics}, Cambridge University 
Press, Cambridge, UK, 1998.

\bibitem{Zwan} R. Zwanzig, Annu. Rev. Phys. Chem., 1965, {\bf 16}, 67.

\bibitem{Mori} H. Mori, Prog. Theor. Phys., 1965, {\bf 33}, 423.

\bibitem{Forster} D.F\"orster, {\em Hydrodynamic Fluctuations, Broken }
{\em Symmetry, and Correlation Functions}, CRC Press, Boca Raton, 2018.

\bibitem{Lutsko} J.F. Lutsko, J. Appl. Phys., 1989, {\bf 65}, 2991.
 
\end{thebibliography}
\end{document}